\documentclass[letterpaper,twocolumn,10pt]{article}

\usepackage{zhanggroup}
\usepackage{tikz}
\usepackage{amsfonts}
\usepackage{xspace} 
\usepackage{amsmath}
\usepackage{mathrsfs}
\usepackage{amssymb}
\usepackage{amsmath}
\usepackage{amsthm}
\usepackage{subcaption}
\usepackage{booktabs}
\hypersetup{
  colorlinks,
  linkcolor={blue!70!green},
  citecolor={green!70!blue},
  urlcolor={orange!70!red}
}

\newtheorem{definition}{Definition}
\newcommand{\mypara}[1]{\smallskip\noindent{\bf {#1}.} \xspace}
\newcommand{\TargetDataset}{\mathcal{D}}
\newcommand{\AdjMatrix}{\mathcal{A}}
\newcommand{\TargetNodeIndexSet}{\mathcal{V}}
\newcommand{\TargetNodeFeature}{\mathcal{F}}
\newcommand{\Knowledge}{\mathcal{K}}
\newcommand{\TargetModel}{f}
\newcommand{\ReferenceModel}{g}
\newcommand{\PartialGraph}{\AdjMatrix^{*}}
\newcommand{\ShadowDataset}{\mathcal{D}^{\prime}}
\newcommand{\ShadowTargetModel}{f^{\prime}}
\newcommand{\ShadowReferenceModel}{g^{\prime}}
\newcommand{\AdjWithSelfLoop}{\tilde{\AdjMatrix}}
\newcommand{\DiagonalMatrix}{\tilde{\mathcal{Q}}}

\newcommand{\Operations}{\Psi}
\begin{document}
% ----------------------------------------------------

%don't want date printed
\date{}

\title{\Large \bf Stealing Links from Graph Neural Networks}

\author{
{\rm Xinlei He\textsuperscript{1}}\ \ \
{\rm Jinyuan Jia\textsuperscript{2}}\ \ \
{\rm Michael Backes\textsuperscript{1}}\ \ \
{\rm Neil Zhenqiang Gong\textsuperscript{2}}\ \ \
{\rm Yang Zhang\textsuperscript{1}}
\\
\\
\textsuperscript{1}\textit{CISPA Helmholtz Center for Information Security}\ \ \ \textsuperscript{2}\textit{Duke University}
}

\maketitle

\pagenumbering{gobble}
\thispagestyle{headings}
\markright{\hfill To appear in the 30th Usenix Security Symposium, August 2021, Vancouver, B.C., Canada\hfill}

% ----------------------------------------------------
\begin{abstract}
% ----------------------------------------------------
Graph data, such as chemical networks and social networks, may be deemed confidential/private because the data owner often spends lots of resources collecting the data or the data contains sensitive information, e.g., social relationships.
Recently, neural networks were extended to graph data, which are known as \emph{graph neural networks (GNNs)}.
Due to their superior performance, GNNs have many applications, such as healthcare analytics, recommender systems, and fraud detection.
In this work, we propose the first attacks to steal a graph from the outputs of a GNN model that is trained on the graph.
Specifically, given a black-box access to a GNN model, our attacks can infer whether there exists a link between any pair of nodes in the graph used to train the model. 
We call our attacks \emph{link stealing attacks}. 
We propose a threat model to systematically characterize an adversary's background knowledge along three dimensions which in total leads to a comprehensive taxonomy of 8 different link stealing attacks.
We propose multiple novel methods to realize these 8 attacks.
Extensive experiments on 8 real-world datasets show that our attacks are effective at stealing links, e.g., AUC (area under the ROC curve) is above 0.95 in multiple cases. 
Our results indicate that the outputs of a GNN model reveal rich information about the structure of the graph used to train the model. 
% ----------------------------------------------------
\end{abstract}
% ----------------------------------------------------

% ----------------------------------------------------
\section{Introduction}
% ----------------------------------------------------

Graph is a powerful tool to model the complex relationships between entities. 
For instance, in healthcare analytics, protein-protein interactions can be modeled as a graph (called a \emph{chemical network}); and a social network can be modeled as a graph, where nodes are users and edges indicate certain social relationships among them. 
A graph may be treated as a data owner's intellectual property because the data owner may spend a lot of resources collecting the graph, e.g., collecting a chemical network often involves expensive and resource-consuming chemical experiments. 
Moreover, a graph may also contain sensitive user information, e.g., private social relationships among users.   

Recently, a family of machine learning techniques known as \emph{graph neural networks (GNNs)} was proposed to analyze graphs.
We consider GNNs for \emph{node classification}. 
Specifically, given a graph, attributes of each node in the graph, and a small number of node labels, a GNN model is trained and can predict the label of each remaining unlabeled node. 
Due to their superior performance, we have seen growing applications of GNNs in various domains, such as healthcare analytics~\cite{GSRVD17,EPBM20}, recommender systems~\cite{FMLHZTY19}, and fraud detection~\cite{WJG19}. 
However, the security and privacy implications of training GNNs on graphs are largely unexplored. 

\mypara{Our Contributions} 
In this work, we take the first step to study the security and privacy implications of training GNNs on graphs. 
In particular, we propose the first attacks to steal a graph from the outputs of a GNN model trained on the graph.  
We call our attacks \emph{link stealing attacks}.
Specifically, given a black-box access to a target GNN model, our attacks aim to predict whether there exists a link between any pair of nodes in the graph used to train the target GNN model. 
Our attacks reveal serious concerns on the intellectual property, confidentiality, and/or privacy of graphs when training GNNs on them.
For instance, our attacks violate the intellectual property of the data owner when it spends lots of resources collecting the graph; and our attacks violate user privacy when the graph contains sensitive social relationships among users~\cite{GL162,BHPZ17}. 

\noindent{\it Adversary's Background Knowledge:} 
We refer to the graph and nodes' attributes used to train the target GNN model as the \emph{target dataset}. 
We characterize an adversary's background knowledge along three dimensions, including the target dataset's \emph{nodes' attributes}, the target dataset's \emph{partial graph}, and an auxiliary dataset (called \emph{shadow dataset}) which also contains its own graph and nodes' attributes.
An adversary may or may not have access to each of the three dimensions. 
Therefore, we obtain a comprehensive taxonomy of a threat model, in which adversaries can have 8 different types of background knowledge. 

\noindent{\it Attack Methodology:} 
We design an attack for each of the 8 different types of background knowledge, i.e., we propose 8 link stealing attacks in total. 
The key intuition of our attacks is that two nodes are more likely to be linked if they share more similar attributes and/or predictions from the target GNN model. 
For instance, when the adversary only has the target dataset's nodes' attributes, we design an unsupervised attack by calculating the distance between two nodes' attributes.
When the target dataset's partial graph is available, we use supervised learning to train a binary classifier as our attack model with features summarized from two nodes' attributes and predictions obtained from the black-box access to the target GNN model. 
When the adversary has a shadow dataset, we propose a \emph{transferring attack} which transfers the knowledge from the shadow dataset to the target dataset to mount our attack.

\noindent{\it Evaluation:} 
We evaluate our 8 attacks using 8 real-world datasets.
First, extensive experiments show that our attacks can effectively steal links. 
In particular, our attacks achieve high AUCs (area under the ROC curve). 
This demonstrates that the predictions of a target GNN model encode rich information about the structure of a graph that is used to train the model, and our attacks can exploit them to steal the graph structure.
Second, we observe that more background knowledge leads to better attack performance in general.
For instance, on the Citeseer dataset~\cite{KW17}, when an adversary has all the three dimensions of the background knowledge, our attack achieves 0.977 AUC.
On the same dataset, when the adversary only has nodes' attributes, the AUC is 0.878.
Third, we find that the three dimensions of background knowledge have different impacts on our attacks. 
Specifically, the target dataset's partial graph has the strongest impact followed by nodes' attributes, the shadow dataset, on the other hand, has the weakest impact.
Fourth, our transferring attack can achieve high AUCs.
Specifically, our transferring attack achieves better performance if the shadow dataset comes from the same domain as the target dataset, e.g., both of them are chemical networks.
We believe this is due to the fact that graphs from the same domain have similar structures, which leads to less information loss during transferring.
Fifth, our attacks outperform conventional link prediction methods~\cite{LK07,GTMHSSSS14}, which aim to predict links between nodes based on a partial graph.  

\smallskip
In summary, we make the following contributions.

\begin{itemize}
\item We propose the first link stealing attacks against graph neural networks.
\item We propose a threat model to comprehensively characterize an adversary's background knowledge along three dimensions. 
Moreover, we propose 8 link stealing attacks for adversaries with different background knowledge.
\item We extensively evaluate our 8 attacks on 8 real-world datasets.  Our results show that our attacks can steal links from a GNN model effectively.  
\end{itemize}

% ----------------------------------------------------
\section{Graph Neural Networks}
\label{section:Gnn}
% ----------------------------------------------------

Many important real-world datasets come in the form of graphs or networks, e.g., social networks, knowledge graph, and chemical networks. 
Therefore, it is urgent to develop machine learning algorithms to fully utilize graph data.
To this end, a new family of machine learning algorithms, i.e., graph neural networks (GNNs), has been proposed and shown superior performance in various tasks~\cite{AT16,DBV16,KW17,VCCRLB18}. 

\mypara{Training a GNN Model} Given a graph,  attributes for each node in the graph, and a small number of labeled nodes, GNN
trains a neural network to predict labels of the remaining unlabeled nodes via analyzing the graph structure and node attributes. 
Formally, we define the \emph{target dataset} as $\TargetDataset = (\AdjMatrix, \TargetNodeFeature)$, where $\AdjMatrix$ is the adjacency matrix of the graph and $\TargetNodeFeature$ contains all nodes' attributes.
Specifically, $\AdjMatrix_{uv}$ is an element in $\AdjMatrix$: If there exists an edge between node $u$ and node $v$, then $\AdjMatrix_{uv}=1$, otherwise $\AdjMatrix_{uv}=0$. 
Moreover, $\TargetNodeFeature_u$ represents the attributes of $u$.
$\TargetNodeIndexSet$ is a set containing all nodes in the graph.
Note that we consider undirected graphs in this paper, i.e., $\forall u, v \in \TargetNodeIndexSet, \AdjMatrix_{uv}=\AdjMatrix_{vu}$.

A GNN method iteratively updates a node's features via aggregating its neighbors' features using a neural network, whose last layer predicts labels for nodes.
Different GNN methods use slightly different aggregation rules. 
For instance, \emph{graph convolutional network (GCN)}, the most representative and well-established GNN method~\cite{KW17}, uses a multi-layer neural network whose architecture is determined by the graph structure. 
Specifically, each layer obeys the following propagation rule to aggregate the neighboring features:
\begin{equation}
\label{equation:GCNlayer}
H^{(k+1)} = \sigma(\DiagonalMatrix^{-\frac{1}{2}} \AdjWithSelfLoop \DiagonalMatrix^{-\frac{1}{2}} H^{(k)}W^{(k)} ),
\end{equation}
where $\AdjWithSelfLoop=\AdjMatrix + I$ is the adjacency matrix of the graph with self-connection added, i.e., $I$ is the identity matrix. 
$\DiagonalMatrix^{-\frac{1}{2}} \AdjWithSelfLoop \DiagonalMatrix^{-\frac{1}{2}}$ is the symmetric normalized adjacency matrix and $\DiagonalMatrix_{uu} = \sum_{u}{\AdjWithSelfLoop_{uv}}$.
Moreover, $W^{(k)}$ is the trainable weight matrix of the $k$th layer and $\sigma(\cdot)$ is the activation function to introduce non-linearity, such as ReLU. 
As the input layer, we have $H^{(0)} = \TargetNodeFeature$. 
When the GCN uses a two-layer neural network, the GCN model can be described as follows:
\begin{equation}
\label{equation:GCNModel}
\textit{softmax}(\DiagonalMatrix^{-\frac{1}{2}} \AdjWithSelfLoop \DiagonalMatrix^{-\frac{1}{2}}\sigma(\DiagonalMatrix^{-\frac{1}{2}} \AdjWithSelfLoop \DiagonalMatrix^{-\frac{1}{2}}\TargetNodeFeature W^{(0)}) W^{(1)}).
\end{equation}
Note that in most of the paper, we focus on two-layer GCN.
Later, we show that our attack can be also performed on other types of GNNs, including GraphSAGE~\cite{HYL17} and GAT~\cite{VCCRLB18} (see \autoref{section:Evaluation}).

\mypara{Prediction in a GNN Model}
Since all nodes' attributes and the whole graph have been fed into the GNN model in the training phase to predict the label of a node, we only need to provide the node's ID to the trained model and obtain the prediction result. 
We assume the prediction result is a posterior distribution (called \emph{posteriors}) over the possible labels for the node. 
Our work shows that such posteriors reveal rich information about the graph structure: As mentioned before, a GNN essentially learns a node's features via aggregating its neighbors' features, if two nodes are connected, then their posteriors should be similar.
We leverage this to build our attack models.
We further use $\TargetModel$ to denote the target GNN model and $\TargetModel(u)$ to represent the posteriors of node $u$.
For presentation purposes, we summarize the notations introduced here and in the following sections in \autoref{table:notation}.

\begin{table}[!t]
\centering
\caption{List of notations.}
\label{table:notation}
\footnotesize
\begin{tabular}{r|l}
\toprule
Notation & Description \\
\midrule
$\TargetDataset$ & Target dataset\\
$\AdjMatrix$ & Graph of $\TargetDataset$ represented as adjacency matrix\\
$\PartialGraph$ & Partial graph of $\TargetDataset$\\
$\TargetNodeFeature$ & Nodes' attributes of $\TargetDataset$\\
$\TargetNodeIndexSet$ & Set of nodes of $\TargetDataset$\\
$\TargetModel$ & Target model\\
$\ReferenceModel$ & Reference model\\
$\TargetModel(u)$ & $u$'s posteriors from the target model\\
$\ReferenceModel(u)$ & $u$'s posteriors from the reference model\\
$\ShadowDataset$ & Shadow dataset\\
$\ShadowTargetModel$ & Shadow target model\\
$\ShadowReferenceModel$ & Shadow reference model\\
$\Knowledge$ & Adversary's knowledge\\
$d(\cdot, \cdot)$ & Distance metric\\
$\Operations(\cdot,\cdot)$ & Pairwise vector operations\\
$e(\TargetModel(u))$ & Entropy of $\TargetModel(u)$\\
\bottomrule
\end{tabular}
\end{table}

% ----------------------------------------------------
\section{Problem Formulation}
\label{section:Problem}
% ----------------------------------------------------

In this section, we first propose a threat model to characterize an adversary's background knowledge.
Then, we formally define our link stealing attack.

% ----------------------------------------------------
\subsection{Threat Model}
% ----------------------------------------------------

\mypara{Adversary's Goal}
An adversary's goal is to infer whether a given pair of nodes $u$ and $v$ are connected in the target dataset. 
Inferring links between nodes leads to a severe privacy threat when the links represent sensitive relationship between users in the context of social networks.
Moreover, links may be confidential and viewed as a model owner's intellectual property because the model owner may spend lots of resources collecting the links, e.g., it requires expensive medical/chemical experiments to determine the interaction/link between two molecules in a chemical network. 
Therefore, inferring links may also compromise a model owner's intellectual property.

\mypara{Adversary's Background Knowledge} 
First, we assume an adversary has a black-box access to the target GNN model. 
In other words, the adversary can only obtain nodes' posteriors by querying the target model $\TargetModel$.
This is the most difficult setting for the adversary~\cite{SSSS17,SZHBFB19,SBBFZ20}.
An adversary can have a black-box access to a GNN model when an organization uses GNN tools from another organization (viewed as an adversary) or the GNN model prediction results are shared among different departments within the same organization. 
For instance, suppose a social network service provider leverages another company’s tool to train a GNN model for fake-account detection, the provider often needs to send the prediction results of (some) nodes to the company for debugging or refining purposes. 
In such a scenario, the security company essentially has a black-box access to the GNN model.  
Note that the graph structure is already revealed to the adversary if she has a white-box access to the target GNN model as the GNN model architecture is often based on the graph structure.  

Then, we characterize an adversary's background knowledge along three dimensions: 

\begin{itemize}
\item \textbf{Target Dataset's Nodes' Attributes, denoted by $\TargetNodeFeature$.} 
This background knowledge characterizes whether the adversary knows  nodes' attributes $\TargetNodeFeature$ in $\TargetDataset$. 
We also assume that the adversary knows labels of a small subset of nodes.
\item \textbf{Target Dataset's Partial Graph, denoted by $\PartialGraph$.} This dimension characterizes whether the adversary knows a subset of links in the target dataset $\TargetDataset$.
Since the goal of link stealing attack is to infer whether there exists an edge/link between a pair of nodes, the partial graph can be used as ground truth edges to train the adversary's attack model. 
\item \textbf{A Shadow Dataset, denoted by $\ShadowDataset$.} This is a dataset which contains its own nodes' attributes and graph.
The adversary can use this to build a GNN model, referred to as \emph{shadow target model} (denoted by $\ShadowTargetModel$) in order to perform a transferring attack.
It is worth noting that the shadow dataset does not need to come from the same domain of the target dataset.
For instance, the shadow dataset can be a chemical network, while the target dataset can be a citation network.
However, results in \autoref{section:Evaluation} show that same-domain shadow dataset indeed leads to better transferring attack performance.
\end{itemize}
We denote the adversary's background knowledge as a triplet:
\[
\Knowledge=(\TargetNodeFeature,\PartialGraph, \ShadowDataset). 
\]
Whether the adversary has each of the three items is a binary choice, i.e., yes or no.
Therefore, we have a comprehensive taxonomy with 8 different types of background knowledge, which leads to 8 different link stealing attacks.
\autoref{table:attack_scenario} summarizes our attack taxonomy.

\begin{table}[!t]
\centering
\caption{
Attack taxonomy. 
$\checkmark$ ($\times$) means the adversary has (does not have) the knowledge.
}
\label{table:attack_scenario}
\footnotesize
\begin{tabular}{l|ccc|l|ccc}
\toprule
Attack & $\TargetNodeFeature$ & $\PartialGraph$ & $\ShadowDataset$ &Attack & $\TargetNodeFeature$ & $\PartialGraph$ & $\ShadowDataset$\\
\midrule
Attack-0 &  $\times$ & $\times$ & $\times$ & Attack-4 & $\times$ &  $\checkmark$&  $\checkmark$\\
Attack-1 &  $\times$&  $\times$& $\checkmark$ & Attack-5 & $\checkmark$ &  $\times$& $\checkmark$\\
Attack-2 &  $\checkmark$& $\times$ &  $\times$ & Attack-6 & $\checkmark$ & $\checkmark$ &  $\times$\\
Attack-3 &  $\times$& $\checkmark$ & $\times$ & Attack-7 & $\checkmark$ & $\checkmark$ & $\checkmark$\\
\bottomrule
\end{tabular}
\end{table}

% ----------------------------------------------------
\subsection{Link Stealing Attack}
% ----------------------------------------------------

After describing our threat model, we can formally define our link stealing attack as follows:

\begin{definition}[Link Stealing Attack]
Given a black-box access to a GNN model that is trained on a target dataset, a pair of nodes $u$ and $v$ in the target dataset, and an adversary's background knowledge $\Knowledge$, link stealing attack aims to infer whether there is a link between $u$ and $v$ in the target dataset.  
\end{definition}

% ----------------------------------------------------
\section{Attack Taxonomy}
\label{section:AttackModels}
% ----------------------------------------------------

\begin{table*}[!t]
\centering
\caption{
Features adopted by our supervised attacks (Attack-3 and Attack 6) and transferring attacks (Attack-1, Attack-4, Attack-5, and Attack-7).
Here, $(\ast)$ means the features are extracted from the shadow dataset in the training phase, and $(\star)$ means the features are extracted from both the shadow dataset and the target dataset (its partial graph) in the training phase.
$d(\cdot, \cdot)$ represents distance metrics defined in \autoref{table:distance}, $\Operations(\cdot, \cdot)$ represents the pairwise vector operations defined in \autoref{table:operator}.
Note that the features used in these attack models include all the distance metrics and pairwise vector operations.
}
\label{table:attack_feature}
\footnotesize
\setlength{\tabcolsep}{0.6em}
\begin{tabular}{l|ccc|ccc|cc}
\toprule
Attack & $d(\TargetModel(u),\TargetModel(v))$ &
$\Operations(\TargetModel(u),\TargetModel(v)))$ &
$\Operations(e(\TargetModel(u)),e(\TargetModel(v)))$ &
$d(\ReferenceModel(u),\ReferenceModel(v))$ &
$\Operations(\ReferenceModel(u),\ReferenceModel(v))$ &
$\Operations(e(\ReferenceModel(u)),e(\ReferenceModel(v)))$ & 
$d(\TargetNodeFeature_u,\TargetNodeFeature_v)$  &  
$\Operations(\TargetNodeFeature_u,\TargetNodeFeature_v)$\\
\midrule
Attack-1 $\ast$ & $\checkmark$ & $\times$ &  $\checkmark$ &  $\times$ &  $\times$ &  $\times$ &  $\times$ &  $\times$ \\
Attack-3 & $\checkmark$ & $\checkmark$ &  $\checkmark$ &  $\times$ &  $\times$ &  $\times$ &  $\times$ &  $\times$ \\
Attack-4 $\star$ & $\checkmark$ & $\times$ &  $\checkmark$ &  $\times$ &  $\times$ &  $\times$ &  $\times$ &  $\times$ \\
Attack-5 $\ast$ & $\checkmark$ & $\times$ &  $\checkmark$ &  $\checkmark$ &  $\times$ &  $\checkmark$ &  $\checkmark$ &  $\times$ \\
Attack-6 & $\checkmark$ & $\checkmark$ &  $\checkmark$ &  $\checkmark$ &  $\checkmark$ &  $\checkmark$ &  $\checkmark$ &  $\checkmark$ \\
Attack-7 $\star$ & $\checkmark$ & $\times$ &  $\checkmark$ &  $\checkmark$ &  $\times$ &  $\checkmark$ &  $\checkmark$ &  $\times$ \\
\bottomrule
\end{tabular}
\end{table*}

In this section, we present the detailed constructions of all the 8 attacks in \autoref{table:attack_scenario}.
Given different knowledge $\Knowledge$, the adversary can conduct their attacks in different ways. 
However, there are two problems that exist across different attacks.

The first problem is \emph{node pair order}. 
As we consider undirected graph, when the adversary wants to predict whether there is a link between two given nodes $u$ and $v$, the output should be the same regardless of the input node pair order.

The second problem is \emph{dimension mismatch}.
The shadow dataset and the target dataset normally have different dimensions with respect to attributes and posteriors (as they are collected for different classification tasks).
For transferring attacks that require the adversary to transfer information from the shadow dataset to the target dataset, it is crucial to keep the attack model's input features' dimension consistent no matter which shadow dataset she has.

We will discuss how to solve these two problems during the description of different attacks.
For presentation purposes, features used in our supervised attacks and transferring attacks are summarised in~\autoref{table:attack_feature}.

% ----------------------------------------------------
\subsection{Attack Methodologies}
\label{subsection:AttackTaxonomy}
% ----------------------------------------------------

\mypara{Attack-0: $\Knowledge=(\times, \times, \times)$}
We start with the most difficult setting for the adversary, that is she has no knowledge of the target dataset's nodes' attributes, partial graph, and a shadow dataset. 
All she has is the posteriors of nodes obtained from the target model $\TargetModel$ (see \autoref{section:Gnn}).

As introduced in \autoref{section:Gnn}, GNN essentially aggregates information for each node from its neighbors.
This means if there is a link between two nodes, then their posteriors obtained from the target model should be closer.
Following this intuition, we propose an unsupervised attack.
More specifically, to predict whether there is a link between $u$ and $v$ , we calculate the distance between their posteriors, i.e., $d(\TargetModel(u), \TargetModel(v))$, as the predictor.

We have in total experimented with 8 common distance metrics: Cosine distance, Euclidean distance, Correlation distance, Chebyshev distance, Braycurtis distance, Canberra distance, Manhattan distance, and Square-euclidean distance. 
Their formal definitions are in~\autoref{table:distance} in Appendix.
It is worth noting that all distance metrics we adopt are symmetric, i.e., $d(\TargetModel(u), \TargetModel(v)) = d(\TargetModel(v), \TargetModel(u))$, this naturally solves the problem of \emph{node pair order}. 

Since the attack is unsupervised, to make a concrete prediction, the adversary needs to manually select a threshold depending on application scenarios.
To evaluate our attack, we mainly use AUC which considers a set of thresholds as previous works~\cite{FLJLPR14,BHPZ17,HZHBTWB19,SZHBFB19,JSBZG19,ZHSMVB20}.
In addition, we propose a threshold estimation method based on clustering (see \autoref{section:Evaluation} for more details). 

\mypara{Attack-1: $\Knowledge=(\times, \times, \ShadowDataset)$} 
In this attack, we broaden the adversary's knowledge with a shadow dataset, i.e., $\ShadowDataset$.
This means the adversary can train a classifier for a supervised attack, more specifically, a \emph{transferring attack}.
She first constructs a shadow target model $\ShadowTargetModel$ with $\ShadowDataset$.
Then, she derives the training data from $\ShadowTargetModel$ to train her attack model. 

The adversary cannot directly use the posteriors obtained from the shadow target model as features to train her attack model, as the shadow dataset and the target dataset very likely have different numbers of labels, i.e., the corresponding posteriors are in different dimensions.
This is the dimension mismatch problem mentioned before.
To tackle this, we need to design features over posteriors.

As discussed in Attack-0, for any dataset, if two nodes are linked, then their posteriors obtained from the target model should be similar.
This means if the attack model can capture the similarity of two nodes' posteriors from the shadow target model, it can transfer the information to the target model.

We take two approaches together to design features.
The first approach is measuring distances between two nodes' posteriors.
To this end, for each pair of nodes $u^{\prime}$ and $v^{\prime}$ from the shadow dataset $\ShadowDataset$, we adopt the same set of 8 metrics used in Attack-0 (formal definitions are listed in~\autoref{table:distance}) to measure their posteriors $\ShadowTargetModel(u^{\prime})$ and $\ShadowTargetModel(v^{\prime})$'s distances,
and concatenate these different distances together.
This leads to an 8-dimension vector.

The second approach is to use entropy to describe each posterior inspired by previous works~\cite{NSH18,JSBZG19}.
Formally, for the posterior of node $u^{\prime}$ obtained from the shadow target model $\ShadowTargetModel$, its entropy is defined as the following.
\begin{equation}
\label{equation:Entropy}
e(\ShadowTargetModel(u^{\prime})) = -\sum_i \ShadowTargetModel_i(u^{\prime}) log(\ShadowTargetModel_i(u^{\prime}))
\end{equation}
where $\ShadowTargetModel_i(u^{\prime})$ denotes the $i$-th element of $\ShadowTargetModel(u^{\prime})$.
Then, for each pair of nodes $u^{\prime}$ and $v^{\prime}$ from the shadow dataset, we obtain two entropies $e(\ShadowTargetModel(u^{\prime}))$ and $e(\ShadowTargetModel(v^{\prime}))$.
To eliminate the node pair order problems for these entropies,
we further take the approach of Grover and Leskovec~\cite{GL16},
by applying pairwise vector operation, denoted by $\Operations(\cdot, \cdot)$.
In total, we have used all the 4 operations defined in \autoref{table:operator} (in Appendix) for our attack.
Note that these operations in \autoref{table:operator} are applied on two single numbers, i.e., scalars, in this attack.
However, they can also be applied to vectors and we will adopt them again on posteriors and nodes' attributes in other attacks.

In total, the features used for training the attack model is assembled with 8 different distances between two nodes' posteriors from the shadow target model and 4 features obtained from pairwise vector operations between two nodes' posteriors' entropies.
Regarding labels for the training set, the adversary uses all the links in $\ShadowDataset$ and samples the same number of node pairs that are not linked (see \autoref{section:Evaluation} for more details).
We adopt an MLP as our attack model.

\mypara{Attack-2: $\Knowledge=(\TargetNodeFeature, \times, \times)$}
In this attack, we assume that the adversary has the knowledge of the target dataset's nodes' attributes $\TargetNodeFeature$. 
Since the adversary has no knowledge of the partial graph and a shadow dataset, her attack here is also unsupervised (similar to Attack-0).
We again rely on the distance metrics to perform our attack.
For each pair of nodes $u$ and $v$ from the target dataset, we consider four types of information to measure distance with all the metrics listed in \autoref{table:distance}.
Similar to Attack-0, we experimentally decide which is the most suitable distance metric for Attack-2.

\begin{itemize}
\item $d(\TargetModel(u), \TargetModel(v))$. 
The first type is the same as the method for Attack-0, i.e., distance between posteriors of $u$ and $v$ from the target model $\TargetModel$, i.e., $\TargetModel(u)$ and $\TargetModel(v)$.
\item $d(\TargetNodeFeature_u, \TargetNodeFeature_v)$. 
The second type is calculating the pairwise distance over $u$ and $v$'s attributes $\TargetNodeFeature_u$ and $\TargetNodeFeature_v$.
\item $d(\TargetModel(u),\TargetModel(v)) - d(\ReferenceModel(u),\ReferenceModel(v))$. 
For the third type, since we have the target model's nodes' attributes (as well as a subset of their corresponding labels), we train a separate MLP model, namely \emph{reference model} (denoted by $\ReferenceModel$). 
Our intuition is that if two nodes are connected, the distance between their posteriors from the target model should be smaller than the corresponding distance from the reference model.
Therefore, we calculate $d(\TargetModel(u),\TargetModel(v)) - d(\ReferenceModel(u),\ReferenceModel(v))$ to make prediction.
\item $d(\ReferenceModel(u), \ReferenceModel(v))$. For the fourth type, we measure the distance over $u$ and $v$'s posteriors from the reference model.
\end{itemize}

\mypara{Attack-3: $\Knowledge=(\times, \PartialGraph, \times)$}
In this scenario, the adversary has access to the partial graph $\PartialGraph$ of the target dataset.
For the attack model, we rely on links from the known partial graph as the ground truth label to train an attack model (we again adopt an MLP).
Features used for Attack-3 are summarized in \autoref{table:attack_feature}.
For each pair of nodes $u$ and $v$ from the target dataset,
we calculate the same set of features proposed for Attack-1 on their posteriors and posteriors' entropies.
Besides, since we can directly train the attack model on the partial target graph (i.e., we do not face the dimension mismatch problem), we further define new features by adopting the pairwise vector operations listed in \autoref{table:operator} to $\TargetModel(u)$ and $\TargetModel(v)$.

\mypara{Attack-4: $\Knowledge=(\times, \PartialGraph, \ShadowDataset)$}
In this attack, the adversary has the knowledge of the partial graph $\PartialGraph$ of the target dataset and a shadow dataset $\ShadowDataset$.
To take both knowledge into consideration, for each pair of nodes either from the shadow dataset or the partial graph of the target dataset, we calculate the same set of features over posteriors as proposed in Attack-1.
This means the only difference between Attack-4 and Attack-1 is that the training set for Attack-4 also includes information from the target dataset's partial graph (see \autoref{table:attack_feature}).

Different from Attack-3, Attack-4 cannot perform the pairwise vector operations to $\TargetModel(u)$ and $\TargetModel(v)$.
This is due to the dimension mismatch problem as the adversary needs to take both $\PartialGraph$ and $\ShadowDataset$ into account for her attack.

\mypara{Attack-5: $\Knowledge=(\TargetNodeFeature, \times, \ShadowDataset)$} 
In this attack, the adversary has the knowledge of the target model's nodes' attributes $\TargetNodeFeature$ and a shadow dataset $\ShadowDataset$.
As we do not have $\PartialGraph$ to train the attack model, we need to rely on the graph of the shadow dataset.
To this end, we first calculate the same set of features used for Attack-1.
Moreover, as we have the target dataset's nodes' attributes, we further build a reference model (as in Attack-2), and also a shadow reference model in order to transfer more knowledge from the shadow dataset for the attack.
For this, we build the same set of features as in Attack-1 over the posteriors obtained from the shadow reference model, i.e., the distance of posteriors (\autoref{table:distance}) and pairwise vector operations performed on posteriors' entropies (\autoref{table:operator}).
In addition, we also calculate the 8 different distances over the shadow dataset's nodes' attributes.

\mypara{Attack-6: $\Knowledge=(\TargetNodeFeature, \PartialGraph, \times)$}
In this scenario, the adversary has the access to the target dataset's nodes' attributes $\TargetNodeFeature$ and the partial target graph $\PartialGraph$. 
As a supervised learning setting, we build an MLP considering links from the partial graph as the ground truth label.
The adversary first adopts the same set of features defined over posteriors obtained from the target model as proposed in Attack-3.
Then, the adversary builds a reference model over the target dataset's nodes' attributes, and calculate the same set of features over posteriors obtained from the reference model.
In the end, we further calculate the distances of the target dataset's nodes' attributes as another set of features.

\mypara{Attack-7: $\Knowledge=(\TargetNodeFeature, \PartialGraph, \ShadowDataset)$}
This is the last attack with the adversary having all three knowledge.
The set of features for this attack is the same as the ones used in Attack-5 (\autoref{table:attack_feature}).
The only difference lies in the training phase, we can use the partial graph from the target dataset together with the graph from the shadow dataset as the ground truth.
We expect this leads to better performance than the one for Attack-5.
However, this attack also relies on the information of the shadow dataset, thus, the features used here are a subset of the ones for Attack-6, this is similar to the difference between Attack-4 and Attack-3.
Note that if the adversary does not take the shadow dataset into consideration, this scenario is equivalent to the one for Attack-6. 

% ----------------------------------------------------
\subsection{Summary}
% ----------------------------------------------------

We propose 8 attack scenarios with the combination of the knowledge that the adversary could have.
They could be divided into three categories.

The first category is unsupervised attacks, i.e., Attack-0 and Attack-2, where the adversary does not have the knowledge about the partial graph from the target dataset or a shadow dataset.
In these scenarios, the adversary can use distance metrics for posteriors or nodes' attributes to infer the link.

The second category is the supervised attacks, including Attack-3 and Attack-6, where the adversary has the knowledge of the partial graph from the target dataset but does not have a shadow dataset.
In these scenarios, the adversary can use different distances and pairwise vector operations over nodes' posteriors (and the corresponding entropies) from the target model and their attributes to build features.

The third category is the transferring attacks (supervised), including Attack-1, Attack-4, Attack-5, and Attack-7, where the adversary has the knowledge of a shadow dataset. 
In these scenarios, the adversary can use distance metrics over posteriors/nodes' attributes and pairwise operations over posteriors' entropies as the bridge to transfer the knowledge from the shadow dataset to perform link stealing attacks.
It is worth noting that for Attack-4 and Attack-7, if the adversary leaves the shadow dataset out of consideration, they will not have the dimension mismatch problem and can take the same attack methods as Attack-3 and Attack-6, respectively.

% ----------------------------------------------------
\section{Evaluation}
\label{section:Evaluation}
% ----------------------------------------------------

This section presents the evaluation results of our 8 attacks.
We first introduce our experimental setup.
Then, we present detailed results for different attacks.
Finally, we summarize our experimental findings. 

% ----------------------------------------------------
\subsection{Experimental Setup}
% ----------------------------------------------------

\mypara{Datasets}
We utilize 8 public datasets, including Citeseer~\cite{KW17}, Cora~\cite{KW17}, Pubmed~\cite{KW17}, AIDS~\cite{RB08}, COX2~\cite{SOW03}, DHFR~\cite{SOW03}, ENZYMES~\cite{DD03}, and PROTEINS\_full~\cite{BOSVSK05}, to conduct our experiments.
These datasets are widely used as benchmark datasets for evaluating GNNs~\cite{KW17,VCCRLB18,DJLBB20,EPBM20}. 
Among them, Citeseer, Cora, and Pubmed are citation datasets with nodes representing publications and links indicating citations among these publications. 
The other five datasets are chemical datasets, each node is a molecule and each link represents the interaction between two molecules. 
All these datasets have nodes' attributes and labels. 

\mypara{Datasets Configuration}
For each dataset, we train a target model and a reference model. 
In particular, we randomly sample 10\% nodes and use their ground truth labels to train the target model and the reference model.\footnote{We do not train the reference model for attacks when $\TargetNodeFeature$ is unavailable.} 
Recall that several attacks require the knowledge of the target dataset's partial graph. 
To simulate and fairly evaluate different attacks, we construct an \emph{attack dataset} which contains node pairs and labels representing whether they are linked or not.
Specifically, we first select all node pairs that are linked.
Then, we randomly sample the same number of node pairs that are not linked.
We note that such negative sampling approach follows the common practice in the literature of link prediction~\cite{GL16,BHPZ17,Z19}.
Furthermore, the main metric we use, i.e., AUC (introduced below), is insensitive to the class imbalance issue~\cite{FLJLPR14,BHPZ17,PZ17} contrary to accuracy.
Next, we split the attack dataset randomly by half into \emph{attack training dataset} and \emph{attack testing dataset}.\footnote{We perform additional experiments and observe that training set size does not have a strong impact on the attack performance, results are presented in \autoref{figure:attack_different_ratio} in Appendix.}
We use the attack training dataset to train our attack models when the target dataset's partial graph is part of the adversary's knowledge. 
We use attack testing dataset to evaluate all our attacks. 
For the attacks that have a shadow dataset, we also construct an attack dataset on the shadow dataset to train the attack model. 
Note that we do not split this attack dataset because we do not use it for evaluation. 

\mypara{Metric}
We use AUC (area under the ROC curve) as our main evaluation metric.
AUC is frequently used in binary classification tasks~\cite{FLJLPR14,BHPZ17,PZ17,PZ172,HZHBTWB19,Z19,JSBZG19}, it is threshold independent.
For convenience, we refer to node pairs that are linked as \emph{positive node pairs} and those that are not linked as \emph{negative node pairs}. 
If we rank node pairs according to the probability that there is a link between them, then AUC is the probability that a randomly selected positive node pair ranks higher than a randomly selected negative node pair. 
When performing random guessing, i.e., we rank all node pairs uniformly at random, the AUC value is 0.5.
Note that we also calculate Precision and Recall for all supervised attacks (see \autoref{table:attack1_pr}, \autoref{table:attack3_pr}, \autoref{table:attack4_pr}, \autoref{table:attack5_pr}, \autoref{table:attack6_pr}, and \autoref{table:attack7_pr} in Appendix).

\mypara{Models}
We use a graph convolutional network with 2 hidden layers for both the target model and the shadow target model, and assume they share the same architecture (see \autoref{section:Problem}). 
Note that we also evaluate the scenario where the target model and the shadow model have different architectures later in this section and find the performances of our attacks are similar.
The number of neurons in the hidden layer is set to 16. 
We adopt the frequently used ReLU and softmax as activation functions for the first hidden layer and the second hidden layer, respectively. 
Note that we append Dropout (the rate is 0.5) to the output of the hidden layer to prevent overfitting. 
We train 100 epochs with a learning rate of 0.01. 
Cross-entropy is adopted as the loss function and we use the Adam optimizer to update the model parameters. 
Our GNNs are implemented based on publicly available code.\footnote{\url{https://github.com/tkipf/gcn}} 
Experimental results show that our GNNs achieve similar performance as reported in other papers.
We omit them to preserve space.

We use an MLP with 2 hidden layers as the reference model and the shadow reference model. 
Hyperparameters, including the number of neurons in the hidden layer, activation functions, loss function, optimizer, epochs, and learning rate are the same as those of the target model. 

\begin{table*}[!t]
\centering
\caption{
Average AUC with standard deviation for Attack-1 on all the 8 datasets.
Best results are highlighted in bold.
}
\label{table:attack1}
\footnotesize
\setlength{\tabcolsep}{0.4em}
\begin{tabular}{l|cccccccc}
\toprule
& \multicolumn{8}{c}{Shadow Dataset}\\
Target Dataset & AIDS & COX2 & DHFR & ENZYMES & PROTEINS\_full & Citeseer & Cora & Pubmed\\
\midrule
AIDS& - &  0.720 $\pm$ 0.009 &  0.690 $\pm$ 0.005 &  \textbf{0.730 $\pm$ 0.010} &  0.720 $\pm$ 0.005 &  0.689 $\pm$ 0.019 &  0.650 $\pm$ 0.025 &  0.667 $\pm$ 0.014\\
COX2 &  0.755 $\pm$ 0.032& - &  0.831 $\pm$ 0.005 &  0.739 $\pm$ 0.116 &  \textbf{0.832 $\pm$ 0.009} &  0.762 $\pm$ 0.009 &  0.773 $\pm$ 0.008 &  0.722 $\pm$ 0.024\\
DHFR &  0.689 $\pm$ 0.004 &  \textbf{0.771 $\pm$ 0.004}& - &  0.577 $\pm$ 0.044 &  0.701 $\pm$ 0.010 &  0.736 $\pm$ 0.005 &  0.740 $\pm$ 0.003 &  0.663 $\pm$ 0.010\\
ENZYMES &  \textbf{0.747 $\pm$ 0.014} &  0.695 $\pm$ 0.023 &  0.514 $\pm$ 0.041& - &  0.691 $\pm$ 0.030 &  0.680 $\pm$ 0.012 &  0.663 $\pm$ 0.009 &  0.637 $\pm$ 0.018\\
PROTEINS\_full &  0.775 $\pm$ 0.020 &  0.821 $\pm$ 0.016 &  0.528 $\pm$ 0.038 &  0.822 $\pm$ 0.020& - &  \textbf{0.823 $\pm$ 0.004} &  0.809 $\pm$ 0.015 &  0.809 $\pm$ 0.013\\
Citeseer &  0.801 $\pm$ 0.040 &  0.920 $\pm$ 0.006 &  0.842 $\pm$ 0.036 &  0.846 $\pm$ 0.042 &  0.848 $\pm$ 0.015& - &  \textbf{0.965 $\pm$ 0.001} &  0.942 $\pm$ 0.003\\
Cora &  0.791 $\pm$ 0.019 &  0.884 $\pm$ 0.005 &  0.811 $\pm$ 0.024 &  0.804 $\pm$ 0.048 &  0.869 $\pm$ 0.012 &  \textbf{0.942 $\pm$ 0.001}& - &  0.917 $\pm$ 0.002\\
Pubmed &  0.705 $\pm$ 0.039 &  0.796 $\pm$ 0.007 &  0.704 $\pm$ 0.042 &  0.708 $\pm$ 0.067 &  0.752 $\pm$ 0.014 &  0.883 $\pm$ 0.006 &  \textbf{0.885 $\pm$ 0.005}& -\\
\bottomrule
\end{tabular}
\end{table*}

We use an MLP with 3 hidden layers as our attack model. 
The number of neurons for all hidden layers is 32. 
ReLU is adopted as the activation function for hidden layers and softmax is used as the output activation function. 
We append Dropout (the rate is 0.5) to each hidden layer to prevent overfitting. 
We train 50 epochs with a learning rate of 0.001. 
The loss function is cross-entropy and the optimizer is Adam. 

We run all experiments with this setting for 5 times and report the average value and the standard deviation of AUC scores. 
Note that for Attack-0 and Attack-2, the AUC scores keep the same since these two attacks are unsupervised.

% ----------------------------------------------------
\subsection{Attack Performance}
% ----------------------------------------------------

\begin{figure}[!t]
\centering
\includegraphics[width=\columnwidth]{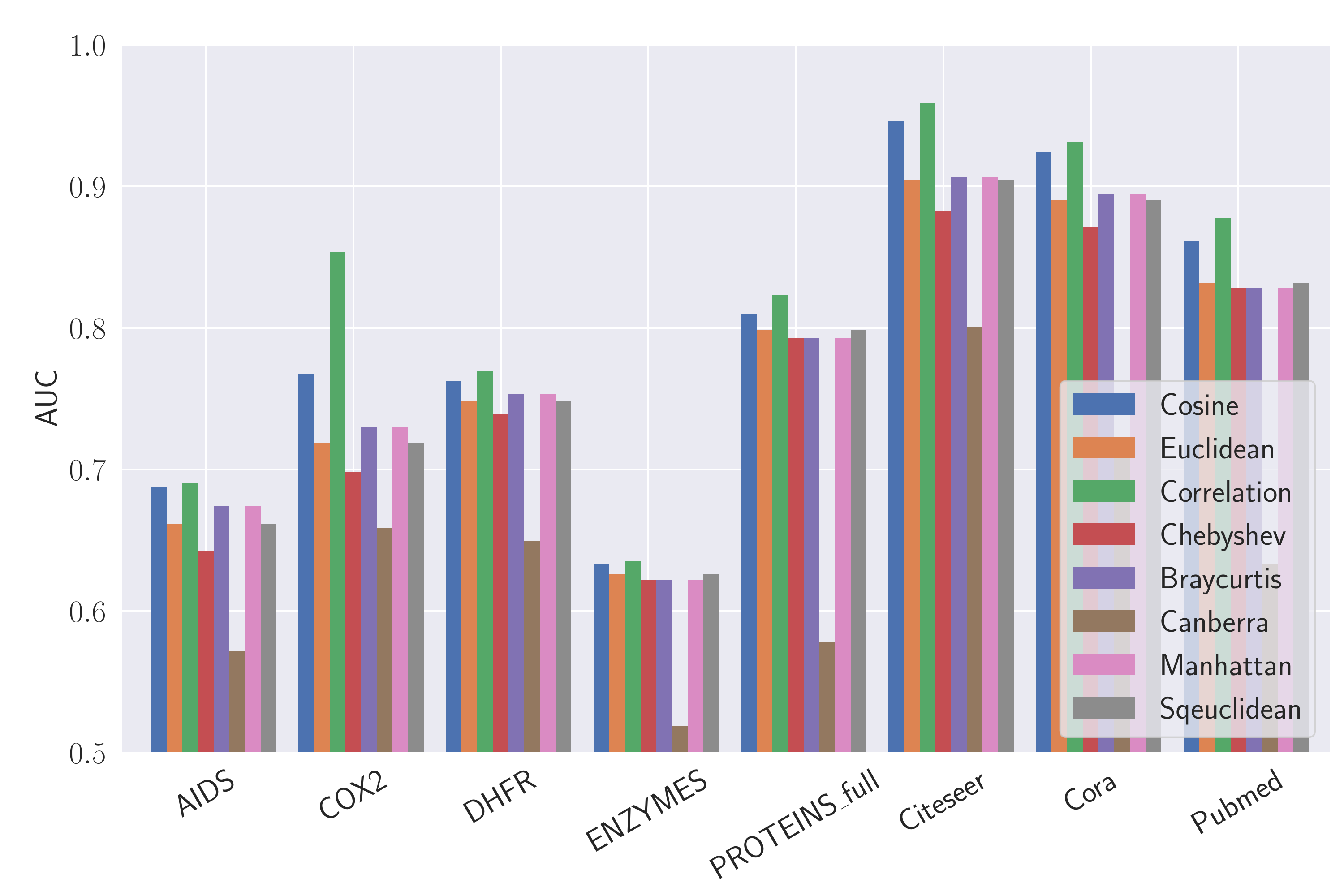}
\caption{
AUC for Attack-0 on all the 8 datasets with all the 8 distance metrics.
The x-axis represents the dataset and the y-axis represents the AUC score.
}
\label{figure:attack0}
\end{figure} 

\begin{figure*}[!t]
\centering
\includegraphics[width=2.0\columnwidth]{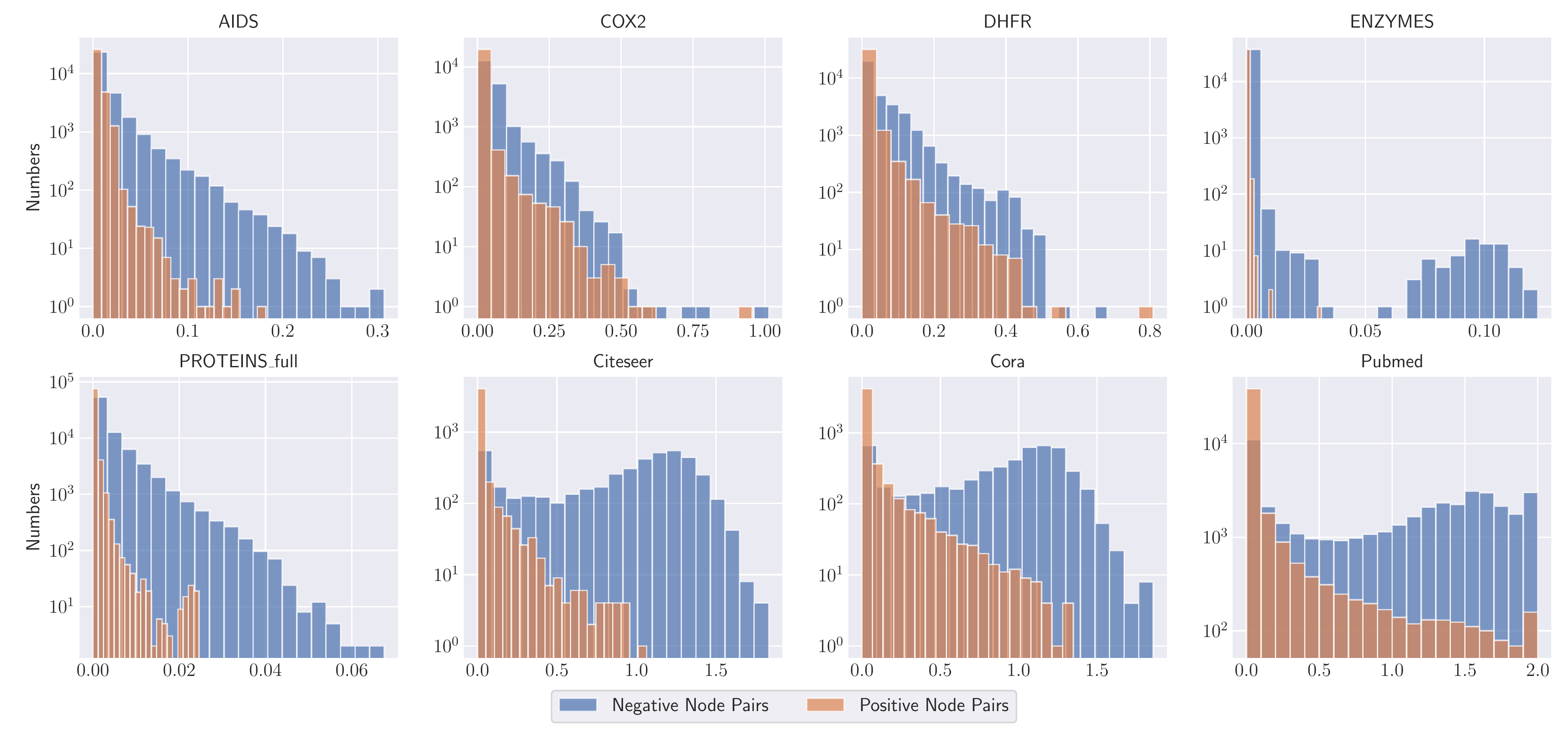}
\caption{
The Correlation distance distribution between nodes' posteriors for positive node pairs and negative node pairs on all the 8 datasets.
The x-axis represents Correlation distance and the y-axis represents the number of node pairs.
}
\label{figure:TargetCorrelation}
\end{figure*} 

\begin{figure*}[!t]
\centering
\begin{subfigure}{1.0\columnwidth}
\centering
\includegraphics[width=1.0\columnwidth]{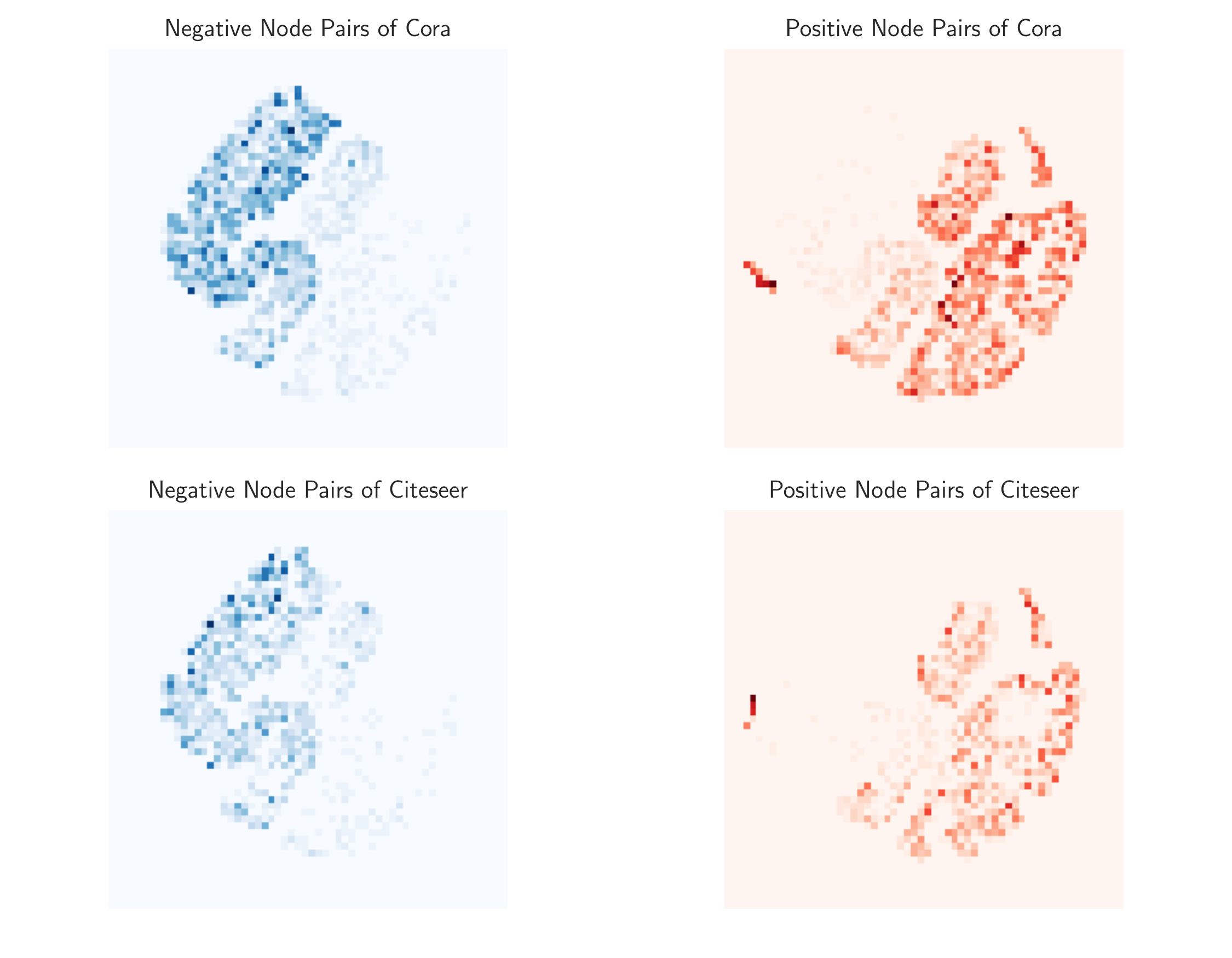}
\caption{}
\label{figure:transfer_cora_citeseer}
\end{subfigure}
\begin{subfigure}{1.0\columnwidth}
\centering
\includegraphics[width=1.0\columnwidth]{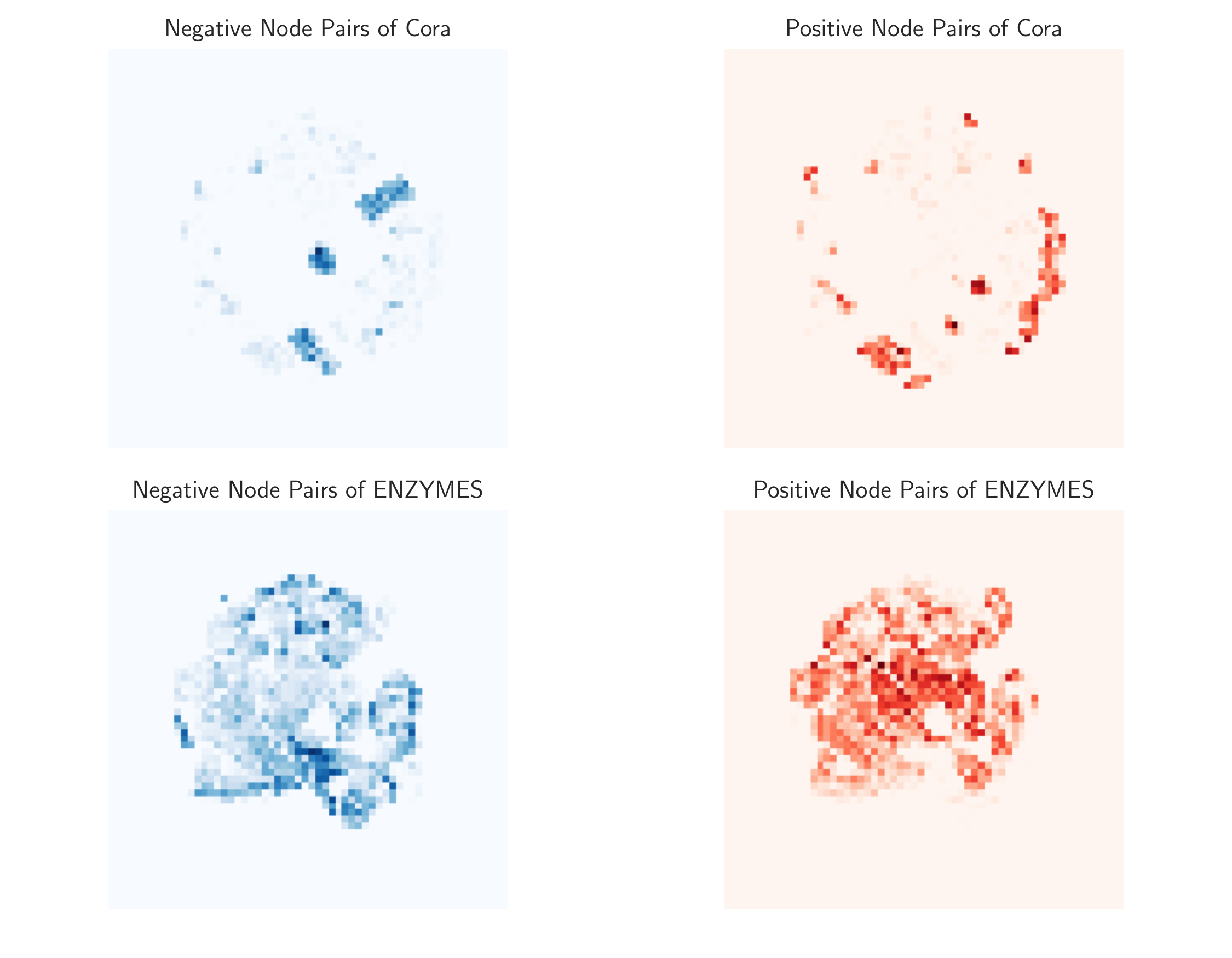}
\caption{}
\label{figure:transfer_cora_ensymes}
\end{subfigure}
\caption{
The last hidden layer's output from the attack model of Attack-1 for 200 randomly sampled positive node pairs and  200 randomly sampled negative node pairs projected into a 2-dimension space using t-SNE. 
(a) Cora as the shadow dataset and Citeseer as the target dataset, 
(b) Cora as the shadow dataset and ENZYMES as the target dataset.
}
\label{figure:transfer_tsne}
\end{figure*} 

\mypara{Attack-0: $\Knowledge=(\times, \times, \times)$} 
In this attack, the adversary only relies on measuring the distance of two nodes' posteriors obtained from the target model. 
We compare 8 different distance metrics and \autoref{figure:attack0} shows the results. 
First, we observe that Correlation distance achieves the best performance followed by Cosine distance across all datasets. 
In contrast, Canberra distance performs the worst. 
For instance, on the Citeseer dataset, the AUC scores for Correlation distance and Cosine distance are 0.959 and 0.946, respectively, while the AUC score for Canberra distance is 0.801.
Note that both Correlation distance and Cosine distance measure the inner product between two vectors, or the ``angle'' of two vectors while other distance metrics do not. 
Second, we find that the performance of the same metric on different datasets is different. 
For instance, the AUC of Correlation distance on Citeseer is 0.959 compared to 0.635 on ENZYMES. 

As mentioned in \autoref{section:AttackModels}, unsupervised attacks could not provide a concrete prediction.
To tackle this, we propose to use clustering, such as K-means.
Concretely, we obtain a set of node pairs' distances, and perform K-means on these distances with K being set to 2.
The cluster with lower (higher) average distance value is considered as the set of positive (negative) node pairs.
Our experiments show that this method is effective.
For instance, on the Citeseer dataset, we obtain 0.788 Precision, 0.991 Recall, and 0.878 F1-Score.
The complete results are summarized in \autoref{table:attack_0_prf} in Appendix. 
Another method we could use is to assume that the adversary has a certain number of labeled edges, either from the target dataset or the shadow dataset.
The former follows the same setting as our Attack-3, Attack-4, Attack-6, and Attack-7, and the latter is equivalent to Attack-1 and Attack-5.
The corresponding results will be shown later.

\autoref{figure:TargetCorrelation} shows the frequency of Correlation distance computed on posteriors obtained from the target model for both positive node pairs and negative node pairs in attack testing datasets.
The x-axis is the value of Correlation distance and the y-axis is the number of pairs.
A clear trend is that for all datasets, the Correlation distance for positive node pairs is much smaller than negative node pairs. 
We select the top 50\% of node pairs with lowest Correlation distance, group them, and calculate the AUC for each group.
Due to the space limit, we only show the result on Pubmed (\autoref{table:attack0_corr}).
We can see that the AUC drops when the Correlation distance increase, which indicates that Attack-0 works better on node pairs with lower Correlation distance.
In general, the posteriors for positive node pairs are ``closer'' than that for negative node pairs. 
This verifies our intuition in \autoref{section:AttackModels}: GNN can be considered as an aggregation function over the neighborhoods, if two nodes are linked, they aggregate with each other's features and therefore become closer.

\begin{table}[!t]
\centering
\caption{
AUC in different Correlation distance levels for Attack-0 on Pubmed. 
}
\label{table:attack0_corr}
\footnotesize
\begin{tabular}{c|c|c|c}
\toprule
Correlation Distance & AUC & Correlation Distance & AUC \\
\midrule
0.00-0.01 & 0.608 & 0.02-0.03 & 0.407 \\
0.01-0.02 & 0.535  & 0.03-0.04 & 0.399 \\

\bottomrule
\end{tabular}
\end{table}

\mypara{Attack-1: $\Knowledge=(\times, \times, \ShadowDataset)$} 
In this attack, the adversary can leverage a shadow dataset.
In particular, for each dataset, we use one of the remaining datasets as the shadow dataset to perform the attack.
\autoref{table:attack1} summarizes the results. 
We leave the blank in the diagonal because we do not use the target dataset itself as its shadow dataset. 

As we can see from \autoref{table:attack1}, the AUC scores from the best-performing shadow dataset have a consistent improvement on almost all datasets compared to Attack-0. 
One exception is the COX2 dataset in which the AUC score decreases by 0.02. 
The results indicate that the adversary can indeed transfer the knowledge from the shadow dataset to enhance her attack. 

An interesting finding is that for a chemical dataset, the best shadow dataset is normally a chemical dataset as well.
Similar results can be observed for citation datasets.
This shows that it is more effective to transfer knowledge across datasets from the same domain.
To better understand this, we extract the attack model's last hidden layer's output (32-dimension) for positive node pairs and negative node pairs and project them into a 2-dimension space using t-Distributed Stochastic Neighbor Embedding (t-SNE)~\cite{MH08}. 
\autoref{figure:transfer_cora_citeseer} shows the results for Citeseer when using Cora as the shadow dataset, both of which are citation datasets. 
We can see that the positive (negative) node pairs from both the target dataset and the shadow dataset can be clustered into similar position, which indicates the positive (negative) node pairs from both datasets have similar distributions.
This means if the attack model learns a decision boundary to separate positive nodes pairs from the negative node pairs on the shadow dataset, this decision boundary can be easily carried over to the target dataset.

In contrast, \autoref{figure:transfer_cora_ensymes} shows the results for ENZYMES (a chemical dataset) when using Cora (a citation dataset) as the shadow dataset. 
We see that the positive (negative) node pairs from the shadow dataset and the target dataset are distributed differently in the 2-dimension space. 
For example, the positive node pairs for Cora are clustered into the outer space of the circle area whereas the positive node pairs for ENZYMES are clustered into the inner space of the circle area.
Therefore, it is hard for the adversary to perform an effective transferring attack.
The underlying reason for this to happen is that graphs from the same domain have analogous graph structures and similar features.
This leads to less information loss for our transferring attack.

\begin{figure}[!t]
\centering
\includegraphics[width=\columnwidth]{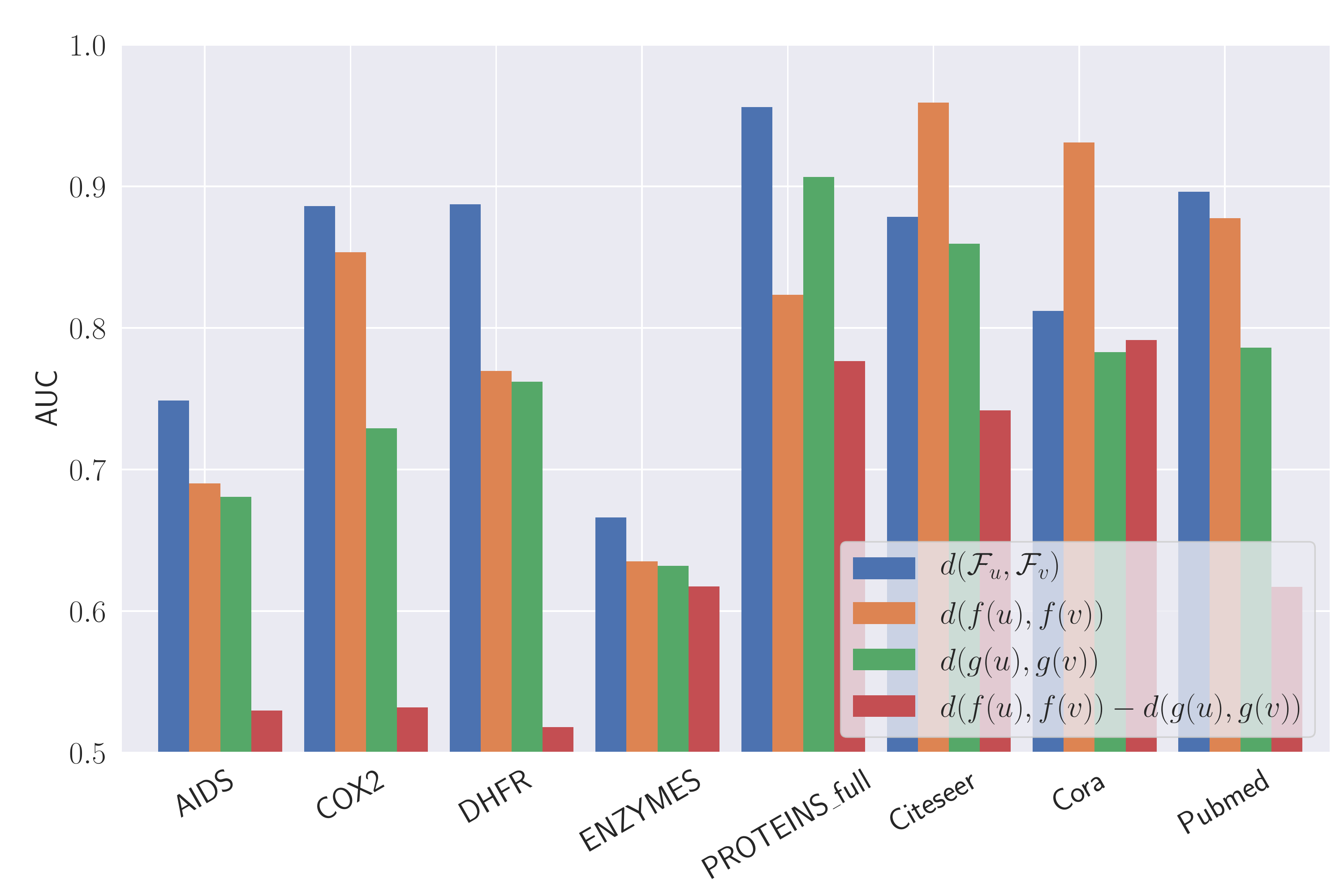}
\caption{
Average AUC for Attack-2 on all the 8 datasets with all the 4 types of information considered.
The x-axis represents the dataset and the y-axis represents the AUC score. 
}
\label{figure:attack2}
\end{figure} 

\mypara{Attack-2: $\Knowledge=(\TargetNodeFeature, \times, \times)$} In Attack-2, the adversary has the knowledge of the target dataset's nodes' attributes.
As discussed in \autoref{section:AttackModels}, she trains a reference model $\ReferenceModel$ by herself from $\TargetNodeFeature$.
We compare four types of information mentioned in \autoref{section:AttackModels}, and the results are shown in \autoref{figure:attack2}.
Note that we only show the results calculated with Correlation distance out of the 8 distance metrics (\autoref{table:distance}) since Correlation distance achieves the best performance in almost all settings.
We can see that in all chemical datasets and one citation dataset, using the distance of target dataset's nodes' attributes leads to the best performance. 
For the other two citation datasets, using the distance between posteriors of the target model can get better performance. 
Nodes' attributes' dimensions are higher in citation datasets than in chemical datasets.
In other words, the node attributes for citation datasets are sparser. 
For instance, we observe that most attributes are 0 in citation datasets.
Therefore, we conclude that the attack can get better performance using the Correlation distance between posteriors of the target model when the target dataset's nodes' attributes are in high dimension.

\begin{table}[!t]
\centering
\caption{
Average AUC with standard deviation for Attack-3 on all the 8 datasets.
}
\label{table:attack3}
\footnotesize
\begin{tabular}{l|c|l|c}
\toprule
Dataset & AUC & Dataset & AUC \\
\midrule
AIDS & 0.961 $\pm$ 0.001 & PROTEINS\_full & 0.958 $\pm$ 0.000\\
COX2 & 0.939 $\pm$ 0.002 & Citeseer & 0.973 $\pm$ 0.000\\
DHFR & 0.934 $\pm$ 0.001 & Cora & 0.954 $\pm$ 0.001\\
ENZYMES & 0.882 $\pm$ 0.001 & Pubmed & 0.947 $\pm$ 0.001\\
\bottomrule
\end{tabular}
\end{table}

\begin{table*}[!t]
\centering
\caption{
Average AUC with standard deviation for Attack-4 on all the 8 datasets.
Best results are highlighted in bold.
}
\label{table:attack4}
\footnotesize
\setlength{\tabcolsep}{0.4em}
\begin{tabular}{l|cccccccc}
\toprule
& \multicolumn{8}{c}{Shadow Dataset}\\
Target Dataset & AIDS & COX2 & DHFR & ENZYMES & PROTEINS\_full & Citeseer & Cora & Pubmed\\
\midrule
AIDS& - &  0.750 $\pm$ 0.009 &  \textbf{0.763 $\pm$ 0.010} &  0.733 $\pm$ 0.007 &  0.557 $\pm$ 0.009 &  0.729 $\pm$ 0.015 &  0.702 $\pm$ 0.010 &  0.673 $\pm$ 0.009\\
COX2 &  0.802 $\pm$ 0.031& - &  \textbf{0.866 $\pm$ 0.004} &  0.782 $\pm$ 0.012 &  0.561 $\pm$ 0.030 &  0.860 $\pm$ 0.002 &  0.853 $\pm$ 0.004 &  0.767 $\pm$ 0.023\\
DHFR &  0.758 $\pm$ 0.022 &  \textbf{0.812 $\pm$ 0.005} & - &  0.662 $\pm$ 0.030 &  0.578 $\pm$ 0.067 &  0.799 $\pm$ 0.002 &  0.798 $\pm$ 0.009 &  0.736 $\pm$ 0.005\\
ENZYMES &  \textbf{0.741 $\pm$ 0.010} &  0.684 $\pm$ 0.024 &  0.670 $\pm$ 0.008& - &  0.733 $\pm$ 0.019 &  0.624 $\pm$ 0.002 &  0.627 $\pm$ 0.014 &  0.691 $\pm$ 0.012\\
PROTEINS\_full &  0.715 $\pm$ 0.009 &  0.802 $\pm$ 0.025 &  0.725 $\pm$ 0.041 &  0.863 $\pm$ 0.010& - &  0.784 $\pm$ 0.031 &  0.815 $\pm$ 0.012 &  \textbf{0.867 $\pm$ 0.003}\\
Citeseer &  0.832 $\pm$ 0.078 &  0.940 $\pm$ 0.005 &  0.914 $\pm$ 0.007 &  0.879 $\pm$ 0.062 &  0.833 $\pm$ 0.088& - &  \textbf{0.967 $\pm$ 0.001} &  0.955 $\pm$ 0.003\\
Cora &  0.572 $\pm$ 0.188 &  0.899 $\pm$ 0.003 &  0.887 $\pm$ 0.014 &  0.878 $\pm$ 0.045 &  0.738 $\pm$ 0.168 &  \textbf{0.945 $\pm$ 0.001} & - &  0.924 $\pm$ 0.005\\
Pubmed &  0.777 $\pm$ 0.056 &  0.893 $\pm$ 0.001 &  0.90 $\pm$ 0.006 &  0.866 $\pm$ 0.002 &  0.806 $\pm$ 0.042 &  \textbf{0.907 $\pm$ 0.004} &  0.902 $\pm$ 0.001& -\\
\bottomrule
\end{tabular}
\end{table*}

\mypara{Attack-3: $\Knowledge=(\times, \PartialGraph, \times)$} 
\autoref{table:attack3} shows the results for this attack. 
With the knowledge of the target dataset's partial graph, the average AUC score for all cases is over 0.9. 
Compared to Attack-2, the AUC scores on chemical datasets have an improvement over 10\% and the AUC scores on citation datasets have an improvement over 2\%.\footnote{Attack-2 achieves relatively high AUC scores on citation datasets.} 

Compared to Attack-1 and Attack-2, Attack-3 achieves the best performance, this indicates the target dataset's partial graph is the most important component for an adversary for performing a link stealing attack.
The reason is that the partial graph contains the ground truth links in the target dataset, which can be directly exploited by the attack model. 

\begin{figure}[!t]
\centering
\includegraphics[width=\columnwidth]{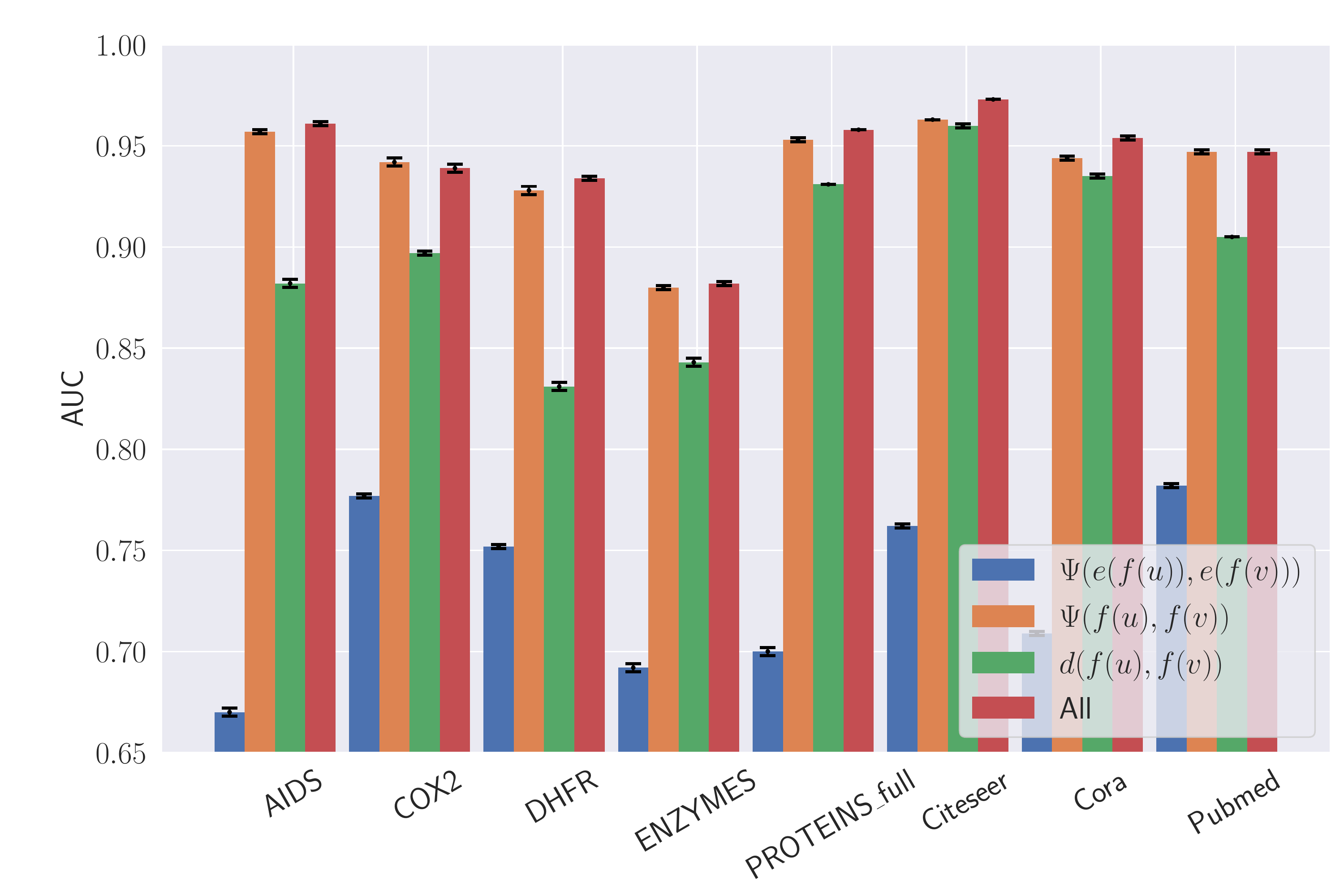}
\caption{
Average AUC for Attack-3 on all the 8 datasets with different set of features. 
The x-axis represents the dataset and the y-axis represents the AUC score. 
}
\label{figure:component}
\end{figure}

We further investigate the contribution of each feature set to the final prediction following the methodology of Dong et al.~\cite{DJC15}. 
Concretely, when studying one feature set, we set other features' value to 0.
As shown in \autoref{figure:component}, the features extracted by applying pairwise operation over posteriors are most useful for the final prediction, followed by the features based on posteriors with different distance metrics. 
We note that our attack also achieves over 0.70 AUC on average when only using pairwise operation over entropy of posteriors as features.
Moreover, our attack achieves the best performance when taking all the three feature sets together, which implies the combination of different features indeed improves the overall performance.

\mypara{Attack-4: $\Knowledge=(\times, \PartialGraph, \ShadowDataset)$}
\autoref{table:attack4} shows the results for Attack-4. 
First, compared to Attack-1 ($\Knowledge=(\times, \times, \ShadowDataset)$), the overall performance of Attack-4 improves with the help of target dataset's partial graph $\PartialGraph$. 
This is reasonable since the target dataset's partial graph contains some ground truth links from the target dataset. 
Second, we note that the performances of Attack-4 are worse than Attack-3 ($\Knowledge=(\times, \PartialGraph, \times)$).
Intuitively, the performance should be better since Attack-4 has more background knowledge. 
The reason for the performance degradation is that we do not take the pairwise vector operation (\autoref{table:operator}) over posteriors as the input for Attack-4 since we want to learn information from both the target dataset and the shadow dataset, and need to eliminate the dimension mismatch issue (as discussed in \autoref{section:AttackModels}).
Moreover, the results also indicate that compared to the shadow dataset, the target dataset's partial graph is more informative.

\mypara{Attack-5: $\Knowledge=(\TargetNodeFeature, \times, \ShadowDataset)$}
In Attack-5, the adversary has the knowledge of target dataset's nodes' attributes as well as a shadow dataset, evaluation results are shown in \autoref{table:attack5}.
We observe that Attack-5 performs better than both Attack-1 (only with $\ShadowDataset$) and Attack-2 (only with $\TargetNodeFeature$).
This shows the combination of $\TargetNodeFeature$ and $\ShadowDataset$ can lead to a better link stealing performance.
Furthermore, we observe similar trends as for Attack-1, that is the attack performs better if the shadow dataset comes from the same domain as the target dataset.

\begin{table*}[!t]
\centering
\caption{
Average AUC with standard deviation for Attack-5 on all the 8 datasets.
Best results are highlighted in bold.
}
\label{table:attack5}
\footnotesize
\setlength{\tabcolsep}{0.4em}
\begin{tabular}{l|cccccccc}
\toprule
& \multicolumn{8}{c}{Shadow Dataset}\\
Target Dataset & AIDS & COX2 & DHFR & ENZYMES & PROTEINS\_full & Citeseer & Cora & Pubmed\\
\midrule
AIDS& - &  0.841 $\pm$ 0.003 &  0.846 $\pm$ 0.009 &  0.795 $\pm$ 0.016 &  \textbf{0.875 $\pm$ 0.002} &  0.839 $\pm$ 0.006 &  0.793 $\pm$ 0.015 &  0.787 $\pm$ 0.008\\
COX2 &  0.832 $\pm$ 0.036& - &  \textbf{0.977 $\pm$ 0.002} &  0.874 $\pm$ 0.020 &  0.946 $\pm$ 0.003 &  0.911 $\pm$ 0.004 &  0.908 $\pm$ 0.004 &  0.887 $\pm$ 0.004\\
DHFR &  0.840 $\pm$ 0.018 &  \textbf{0.988 $\pm$ 0.001}& - &  0.757 $\pm$ 0.032 &  0.970 $\pm$ 0.004 &  0.909 $\pm$ 0.010 &  0.911 $\pm$ 0.009 &  0.860 $\pm$ 0.004\\
ENZYMES &  0.639 $\pm$ 0.005 &  0.581 $\pm$ 0.010 &  0.587 $\pm$ 0.005& - &  0.608 $\pm$ 0.001 &  \textbf{0.685 $\pm$ 0.005} &  0.674 $\pm$ 0.007 &  0.663 $\pm$ 0.002\\
PROTEINS\_full &  0.948 $\pm$ 0.007 &  \textbf{0.981 $\pm$ 0.004} &  0.968 $\pm$ 0.014 &  0.818 $\pm$ 0.017& - &  0.970 $\pm$ 0.002 &  0.876 $\pm$ 0.010 &  0.885 $\pm$ 0.003\\
Citeseer &  0.773 $\pm$ 0.048 &  0.666 $\pm$ 0.018 &  0.652 $\pm$ 0.020 &  0.860 $\pm$ 0.049 &  0.794 $\pm$ 0.009& - &  \textbf{0.969 $\pm$ 0.002} &  0.967 $\pm$ 0.001\\
Cora &  0.743 $\pm$ 0.017 &  0.587 $\pm$ 0.012 &  0.568 $\pm$ 0.009 &  0.778 $\pm$ 0.052 &  0.686 $\pm$ 0.018 &  \textbf{0.956 $\pm$ 0.001}& - &  0.936 $\pm$ 0.002\\
Pubmed &  0.777 $\pm$ 0.030 &  0.661 $\pm$ 0.018 &  0.645 $\pm$ 0.008 &  0.786 $\pm$ 0.041 &  0.741 $\pm$ 0.008 &  0.938 $\pm$ 0.007 &  \textbf{0.941 $\pm$ 0.007}& -\\
\bottomrule
\end{tabular}
\end{table*}

\begin{table*}[!ht]
\centering
\caption{
Average AUC with standard deviation for Attack-7 on all the 8 datasets.
Best results are highlighted in bold.
}
\label{table:attack7}
\footnotesize
\setlength{\tabcolsep}{0.4em}
\begin{tabular}{l|cccccccc}
\toprule
& \multicolumn{8}{c}{Shadow Dataset}\\
Target Dataset & AIDS & COX2 & DHFR & ENZYMES & PROTEINS\_full & Citeseer & Cora & Pubmed\\
\midrule
AIDS& - &  \textbf{0.925 $\pm$ 0.001} &  0.913 $\pm$ 0.005 &  0.784 $\pm$ 0.010 &  0.848 $\pm$ 0.010 &  0.538 $\pm$ 0.022 &  0.520 $\pm$ 0.011 &  0.849 $\pm$ 0.004\\
COX2 &  0.954 $\pm$ 0.007& - &  \textbf{0.982 $\pm$ 0.001} &  0.874 $\pm$ 0.010 &  0.898 $\pm$ 0.030 &  0.947 $\pm$ 0.003 &  0.940 $\pm$ 0.007 &  0.875 $\pm$ 0.034\\
DHFR &  0.982 $\pm$ 0.002 &  \textbf{0.992 $\pm$ 0.00}& - &  0.871 $\pm$ 0.017 &  0.966 $\pm$ 0.008 &  0.933 $\pm$ 0.008 &  0.947 $\pm$ 0.012 &  0.937 $\pm$ 0.003\\
ENZYMES &  \textbf{0.698 $\pm$ 0.007} &  0.691 $\pm$ 0.008 &  0.671 $\pm$ 0.003& - &  0.610 $\pm$ 0.001 &  0.657 $\pm$ 0.009 &  0.662 $\pm$ 0.006 &  0.677 $\pm$ 0.001\\
PROTEINS\_full &  0.984 $\pm$ 0.002 &  0.962 $\pm$ 0.010 &  0.986 $\pm$ 0.002 &  \textbf{0.993 $\pm$ 0.001}& - &  0.840 $\pm$ 0.013 &  0.823 $\pm$ 0.006 &  0.987 $\pm$ 0.005\\
Citeseer &  0.816 $\pm$ 0.048 &  0.791 $\pm$ 0.033 &  0.702 $\pm$ 0.025 &  0.880 $\pm$ 0.057 &  0.902 $\pm$ 0.026& - &  \textbf{0.977 $\pm$ 0.000} &  0.964 $\pm$ 0.000\\
Cora &  0.746 $\pm$ 0.068 &  0.680 $\pm$ 0.038 &  0.574 $\pm$ 0.038 &  0.888 $\pm$ 0.014 &  0.695 $\pm$ 0.10 &  \textbf{0.960 $\pm$ 0.001}& - &  0.935 $\pm$ 0.001\\
Pubmed &  0.807 $\pm$ 0.016 &  0.712 $\pm$ 0.025 &  0.710 $\pm$ 0.006 &  0.881 $\pm$ 0.009 &  0.739 $\pm$ 0.012 &  \textbf{0.956 $\pm$ 0.001} &  0.949 $\pm$ 0.001& -\\
\bottomrule
\end{tabular}
\end{table*}

\begin{table}[!ht]
\centering
\caption{
Average AUC with standard deviation for Attack-6 on all the 8 datasets.
}
\label{table:attack6}
\footnotesize
\begin{tabular}{l|c|l|c}
\toprule
Dataset & AUC & Dataset & AUC \\
\midrule
AIDS & 0.979 $\pm$ 0.001 & PROTEINS\_full & 0.999 $\pm$ 0.000\\
COX2 & 0.987 $\pm$ 0.001 & Citeseer & 0.981 $\pm$ 0.000\\
DHFR & 0.992 $\pm$ 0.001 & Cora & 0.964 $\pm$ 0.000\\
ENZYMES & 0.891 $\pm$ 0.001 & Pubmed & 0.970 $\pm$ 0.000\\
\bottomrule
\end{tabular}
\end{table}

\mypara{Attack-6: $\Knowledge=(\TargetNodeFeature, \PartialGraph, \times)$}
The result of Attack-6 on all datasets is shown in \autoref{table:attack6}. 
We can see that for almost all datasets (except ENZYMES), the AUC scores are over 0.95, which means this attack achieves an excellent performance.
In particular, the AUC score is nearly 1 on PROTEINS\_full. 
Moreover, Attack-6 consistently outperforms Attack-2 ($\Knowledge=(\TargetNodeFeature, \times, \times)$).
This further validates the effectiveness of $\PartialGraph$ in helping the adversary to infer links.  
Another finding is that for chemical datasets, the information of target dataset's partial graph brings a larger improvement than the citation datasets. 
One possible explanation is that the nodes' attributes in chemical datasets contain less information (they are in lower dimension), thus the target dataset's partial graph contributes more to the final prediction performance.

\mypara{Attack-7: $\Knowledge=(\TargetNodeFeature, \PartialGraph, \ShadowDataset)$}
The results of Attack-7 are summarized in \autoref{table:attack7}. 
Compared to Attack-5 ($\Knowledge=(\TargetNodeFeature, \times,  \ShadowDataset)$), the overall performances improve with the help of $\PartialGraph$. 
We would expect the adversary's accuracy is better than that of Attack-6 ($\Knowledge=(\TargetNodeFeature, \PartialGraph, \times)$) since she has more background knowledge.
However, we observe that the performance drops from Attack-6 to Attack-7.
We suspect this is due to the fact that we want to learn information from both the target dataset and the shadow dataset, to avoid the dimension mismatch problem, Attack-7 uses fewer features than Attack-6 (similar to the reason that Attack-4 performs worse than Attack-3).

\begin{figure}[!t]
\centering
\includegraphics[width=\columnwidth]{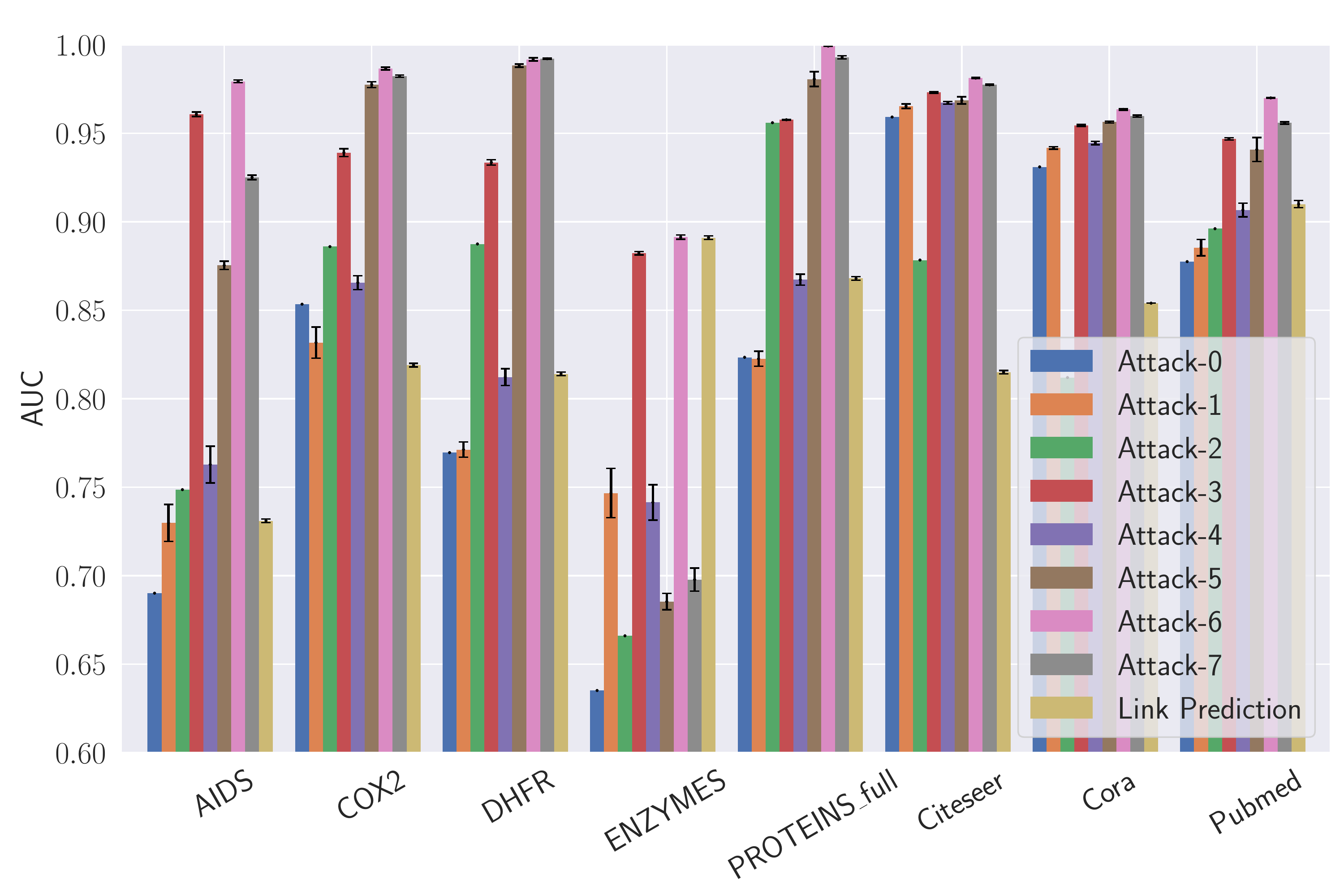}
\caption{
Average AUC with standard deviation for all the attacks on all the 8 datasets.
For each attack, we list its best result.
The x-axis represents the dataset and the y-axis represents the AUC score.
}
\label{figure:attack_all}
\end{figure} 

\mypara{Comparison with Link Prediction}
We further compare all our attacks with a traditional link prediction method~\cite{LK07}.
More specifically, we build an MLP with features summarized from the target model's partial graph, including Common neighbor, Jaccard index, and Preferential attachment~\cite{LK07}. 
As we can see from \autoref{figure:attack_all}, most of our attacks outperforms the link prediction method.
For instance, on the COX2 dataset, all our 8 attacks outperform the link prediction model, the best attack (Attack-6) achieves more than 20\% performance gain.
This demonstrates that GNNs lead to more severe privacy risks than traditional link prediction.

\mypara{Effect of Different GNN Structures}
In our experiments, we adopt the same architecture for both the target model and the shadow target model by default for transferring attack scenarios.
We further evaluate the impact of the shadow target model using different architectures. 
Note that for space reasons, we only report the results of Attack-1. 
Results for other attacks are similar.
We set the number of hidden layers to 3 for the shadow target model (the target model has 2 hidden layers). 
The results are summarized in \autoref{table:attack1_different_layer} in Appendix. 
We find the average AUC scores of our attack are maintained at the same level or even higher for certain datasets compared with the scenario where the shadow target model and the shadow model have the same architecture.
For instance, on the Citeseer dataset, we obtain 0.924 AUC, while the original attack achieves 0.965.
In other words, our attacks are still effective when the shadow target model and the target model have different architectures.

\mypara{Attacks on Other GNNs}
We further investigate whether our attacks are applicable to other GNN models besides GCN.
Concretely, we focus on GraphSAGE~\cite{HYL17} and GAT~\cite{VCCRLB18}. 
We implement GraphSAGE\footnote{\url{https://github.com/williamleif/GraphSAGE}} and GAT\footnote{\url{https://github.com/PetarV-/GAT}} based on publicly available code and only report the results of Attack-6.
\autoref{table:attack6_GraphSAGE} shows that our attack has similar AUC scores on GraphSAGE and GAT compared to GCN. 
For instance, on the COX2 dataset, our attack against GraphSAGE and GAT achieves AUC of 0.982 and 0.984, respectively (the corresponding AUC for GCN is 0.987).
This further demonstrates that our attacks are generally applicable.

\begin{table}[!t]
\centering
\caption{
Average AUC with standard deviation for Attack-6 when using GraphSAGE or GAT as the target model on all the 8 datasets.
}
\label{table:attack6_GraphSAGE}
\footnotesize
\begin{tabular}{l|c|c}
\toprule
Dataset & AUC (GraphSAGE) & AUC (GAT)\\
\midrule
AIDS & 0.977 $\pm$ 0.002  & 0.968 $\pm$ 0.001 \\
COX2 & 0.982 $\pm$ 0.001  & 0.984 $\pm$ 0.001 \\
DHFR & 0.990 $\pm$ 0.001  & 0.995 $\pm$ 0.000 \\
ENZYMES & 0.747 $\pm$ 0.001 & 0.766 $\pm$ 0.004 \\
PROTEINS\_full & 0.999 $\pm$ 0.000 & 0.999 $\pm$ 0.000 \\
Citeseer & 0.938 $\pm$ 0.000 & 0.972 $\pm$ 0.000 \\
Cora & 0.883 $\pm$ 0.001 & 0.958 $\pm$ 0.000 \\
Pubmed & 0.923 $\pm$ 0.000 & 0.965 $\pm$ 0.000 \\
\bottomrule
\end{tabular}
\end{table}

\mypara{Possible Defense}
We try to restrict the GNN model to output $k$ largest posteriors as a defense mechanism to mitigate our attacks. 
The intuition is that the smaller $k$ is, the less information the model reveals.
Here, we fix $k=2$ and report the results for Attack-3. 
Note that we have similar observations for other attacks. 
Experimental results in \autoref{table:attack3_top2} show that this defense indeed reduces the performance of our attack. 
However, the performance drop is not very big, i.e., our attack still achieves relatively high AUC scores. 
For instance, on the Citeseer dataset, this defense reduces Attack-3's performance by less than 2\%.
On the AIDS dataset, the attack's performance drop is higher but AUC being 0.855 still indicates our attack is effective.
We also note that the defense will impact the utility of the model. 
In other words, it is a trade-off between utility and privacy. 
In conclusion, the top-$k$ defense is not effective enough to defend against our attacks.

We can also leverage differential privacy (DP) and adversarial examples to mitigate our attacks. 
In detail, we can adopt edge-DP developed for social networks~\cite{HLMJ09,ZCPSX15} to defend against our attacks.
Borrowing the idea from previous work~\cite{JG18,JSBZG19}, we can also add carefully crafted noise to the prediction of GNN to fool the adversary.
We plan to explore both of them in the future.

\begin{table}[!t]
\centering
\caption{
Average AUC with standard deviation for Attack-3 when only reporting top-2 posteriors on all the 8 datasets.
}
\label{table:attack3_top2}
\footnotesize
\begin{tabular}{l|c|l|c}
\toprule
Dataset & AUC & Dataset & AUC \\
\midrule
AIDS & 0.855 $\pm$ 0.004 & PROTEINS\_full & 0.954 $\pm$ 0.001\\
COX2 & 0.839 $\pm$ 0.005 & Citeseer & 0.958 $\pm$ 0.000\\
DHFR & 0.851 $\pm$ 0.003 & Cora & 0.945 $\pm$ 0.001\\
ENZYMES & 0.876 $\pm$ 0.002 & Pubmed & 0.946 $\pm$ 0.001\\
\bottomrule
\end{tabular}
\end{table}

\mypara{Summary of Results}
In summary, we have made the following observations from our experimental results.

\begin{itemize}
\item Our attacks can effectively steal the links from GNNs. 
For instance, our Attack-6 can achieve average AUC scores over 0.95 on 7 out of 8 datasets, which demonstrate that the GNNs are vulnerable to our attacks. 
\item Generally speaking, the performances of the attack are better if there is more background knowledge as shown in \autoref{figure:attack_all}.
However, we find the impact of different knowledge is different. 
In particular, the target dataset's partial graph is the most informative. 
For instance, Attack-3 ($\Knowledge=(\times, \PartialGraph, \times)$) significantly outperforms Attack-1 ($\Knowledge=(\times, \times, \ShadowDataset)$) and Attack-2 ($\Knowledge=(\TargetNodeFeature, \times, \times)$). 
\item Our transferring attack can achieve good performance. 
Furthermore, we find that our transferring attack achieves better performance when the shadow dataset and the target dataset are from the same domain as validated by experimental results for Attack-1 and Attack-5. 
\end{itemize}

% ----------------------------------------------------
\section{Related Work}
\label{section:RelatedWork}
% ----------------------------------------------------

Various research has shown that machine learning models are vulnerable to security and privacy attacks~\cite{TZJRR16,PMGJCS17,SHNSSDG18,LF20,QMR19,LHZG19,JCBKP20,CWSJ20,SWBMZ20,LMXZ20,CZWBHZ20}.
In this section, we mainly survey four of these attacks that are most relevant to ours.

\mypara{Membership Inference}
In membership inference attacks~\cite{SSSS17,SZHBFB19,NSH18,YGFJ18,HMDC19,NSH19,CXXLBKZ18,SS202,CLEKS19,LZ20}, the adversary aims to infer whether a data sample is in the target model's training dataset or not. 
Shokri et al.~\cite{SSSS17} propose the first membership inference attacks against machine learning models and demonstrate its relationship with model overfitting. 
Salem et al.~\cite{SZHBFB19} further show membership inference attacks are broadly applicable at low cost via relaxing assumptions on the adversary. 
To mitigate attacks, many empirical defenses~\cite{SSSS17,SZHBFB19,NSH18,JSBZG19} have been proposed. 
For instance, Nasr et al.~\cite{NSH18} propose to mitigate attacks via formulating the defense as a min-max optimization problem which tries to decrease the accuracy loss and increase the membership privacy. 
Salem et al.~\cite{SZHBFB19} explore dropout and model stacking to mitigate membership inference attacks. 
More recently, Jia et al.~\cite{JSBZG19} leverage adversarial examples to fool the adversary and show their defense has a formal utility guarantee. 
Other attacks in this space study membership inference in natural language processing models~\cite{SS19}, generative models~\cite{HMDC19,CYZF20}, federated learning~\cite{MSCS19}, and biomedical data~\cite{HZHBTWB19}.

\mypara{Model Inversion} 
In model inversion attacks~\cite{FLJLPR14,FJR15,PMSW18,MSCS19,JCBKP20}, the adversary aims to learn sensitive attributes of training data from target models.
For example, Fredrikson et al.~\cite{FLJLPR14} propose the model inversion attack in which the adversary can infer the patient's genetic markers given the model and some demographic information about the patients. 
Fredrikson et al.~\cite{FJR15} further explore the model inversion attacks on decision trees and neural networks via exploiting the confidence score values revealed along with predictions. 
Melis et al.~\cite{MSCS19} revealed that in the collaborative learning scenarios, when the target model updated with new training data, the adversary could infer sensitive attributes about the new training data.

\mypara{Model Extraction} 
In model extraction attacks~\cite{TZJRR16,WG18,JCBKP20,CCGJY20}, the adversary aims to steal the parameters of a certain target model or mimic its behaviors. 
Tram\'{e}r et al.~\cite{TZJRR16} show that an adversary can exactly recover the target model's parameters via solving the equations for certain models, e.g., linear models. 
Wang and Gong~\cite{WG18} propose attacks to steal the hyperparameters and show their attacks are broadly applicable to a variety of machine learning algorithms, e.g., ridge regression and SVM. 
Orekondy et al.~\cite{OSF19} propose a functionality stealing attack aiming at mimicking the behaviors of the target model.
Concretely, they query the target model and use the query-prediction pairs to train a ``knockoff'' model. 
Jagielski et al.~\cite{JCBKP20} improve the query efficiency of learning-based model extraction attacks and develop the practical functionally-equivalent model whose predictions are identical to the target model on all inputs without training model's weights.
Some defenses~\cite{JSMA19,OSF20} have been proposed to defend against model extraction attacks. 
For instance, Juuti et al.~\cite{JSMA19} propose to detect malicious queries via analyzing the distribution of consecutive API queries and raises an alarm when the distribution different from benign queries. 
Orekondy et al~\cite{OSF20} propose a utility-constrained defense against neural network model stealing attacks via adding perturbations to the output of the target model. 

\mypara{Adversarial Attacks on Graph Neural Networks}
Some recent studies~\cite{ZAG18,BG192,DLTHWZS18,ZG19,WWTDLZ19,WG19,ZJWG20} show that GNNs are vulnerable to adversarial attacks. 
In particular, the adversary can fool GNNs via manipulating the graph structure and/or node features. 
For instance, Z{\"{u}}gner et al.~\cite{ZAG18} introduce adversarial attacks to attributed graphs and focus on both training and testing phase. In particular, their attacks target both node's features and graph structure and show that the node classification accuracy drops with a few perturbations.  
Bojchevski et al.~\cite{BG192} analyze the vulnerability of node embeddings to graph structure perturbation via solving a bi-level optimization problem based on eigenvalue perturbation theory.
Z{\"{u}}gner and G{\"{u}}nnemann~\cite{ZG19} investigate training time attacks on GNNs for node classification via treating the graph as a hyperparameter to optimize. 
Wang and Gong~\cite{WG19} propose an attack to evade the collective classification based classifier via perturbing the graph structure, which can also transfer to GNNs. 
Dai et al.~\cite{DLTHWZS18} propose to fool the GNNs via manipulating the combinatorial structure of data and try to learn generalizable attack policy via reinforcement learning. 
Zhang et al.~\cite{ZJWG20} propose a subgraph based backdoor attack to GNN based graph classification. 
In particular, a GNN classifier outputs a target label specified by an adversary when a predefined subgraph is injected to the testing graph.
These studies are different from our work since we aim to steal links from GNNs. 

To mitigate attacks, many defenses~\cite{BG19,ZZCZ19,WWTDLZ19,ZG192} have been proposed. 
For instance, Zhu et al.~\cite{ZZCZ19} propose to enhance the robustness of GCNs via using Gaussian distributions in graph convolutional layers to mitigate the effects of adversarial attacks and leveraged attention mechanism to impede the propagation of attacks. 
Z{\"{u}}gner and G{\"{u}}nnemann~\cite{ZG192} propose a learning principle that improves the robustness of the GNNs and show provable robustness guarantees against nodes' attributes perturbation. 
Bojchevski et al.~\cite{BG192} propose to certify the robustness against graph structure perturbation for a general class of models, e.g., GNNs, via exploiting connections to PageRank and Markov decision processes. 
These defenses are designed to improve the robustness of GNNs rather than preventing the privacy leakage of it. 
Note that there are also some attacks and defenses on graph that focus on non-GNN models~\cite{CNKMPAV17,JWCG20}. 
For instance, Chen et al.~\cite{CNKMPAV17} propose attacks that mislead the behavior of graph-cluster algorithm and show some practical defenses. 
Jia et al.~\cite{JWCG20} propose certified defense which is based on randomized smoothing to defend against adversarial structural attacks to community detection. 

% ----------------------------------------------------
\section{Conclusion and Future Work}
\label{section:Conclusion}
% ----------------------------------------------------

In this paper, we propose the first link stealing attacks against GNNs.
Specifically, we show that, given a black-box access to a target GNN model, an adversary can accurately infer whether there exists a link between any pair of nodes in a graph that is used to train the GNN model. 
We propose a threat model to systematically characterize an adversary's background knowledge along three dimensions. By jointly considering the three dimensions, we define 8 link stealing attacks and propose novel methods to realize them.
Extensive evaluation over 8 real-world datasets shows that our attacks can accurately steal links. 
Interesting future work includes generalizing our attacks to GNNs for graph classification and defending against our attacks. 

% ----------------------------------------------------
\section*{Acknowledgments}
% ----------------------------------------------------

We thank the anonymous reviewers and our shepherd Minhui Xue for constructive feedback. 
This work is partially funded by the Helmholtz Association within the project ``Trustworthy Federated Data Analytics'' (TFDA) (funding number ZT-I-OO1 4) and National Science Foundation grant No.\ 1937787. 

% ----------------------------------------------------
\bibliographystyle{plain}
\bibliography{normal_generated_py3}

\begin{thebibliography}{10}

\bibitem{AT16}
James Atwood and Don Towsley.
\newblock {Diffusion-Convolutional Neural Networks}.
\newblock In {\em {Annual Conference on Neural Information Processing Systems
  (NIPS)}}, pages 1993--2001. NIPS, 2016.

\bibitem{BHPZ17}
Michael Backes, Mathias Humbert, Jun Pang, and Yang Zhang.
\newblock {walk2friends: Inferring Social Links from Mobility Profiles}.
\newblock In {\em {ACM SIGSAC Conference on Computer and Communications
  Security (CCS)}}, pages 1943--1957. ACM, 2017.

\bibitem{BG192}
Aleksandar Bojchevski and Stephan G{\"u}nnemann.
\newblock {Adversarial Attacks on Node Embeddings via Graph Poisoning}.
\newblock In {\em {International Conference on Machine Learning (ICML)}}, pages
  695--704. PMLR, 2019.

\bibitem{BG19}
Aleksandar Bojchevski and Stephan G{\"u}nnemann.
\newblock {Certifiable Robustness to Graph Perturbations}.
\newblock In {\em {Annual Conference on Neural Information Processing Systems
  (NeurIPS)}}, pages 8317--8328. NeurIPS, 2019.

\bibitem{BOSVSK05}
Karsten~M. Borgwardt, Cheng~Soon Ong, Stefan Schönauer, S.~V.~N. Vishwanathan,
  Alexander~J. Smola, and Hans-Peter Kriegel.
\newblock {Protein Function Prediction via Graph Kernels}.
\newblock {\em {Bioinformatics}}, 2005.

\bibitem{CLEKS19}
Nicholas Carlini, Chang Liu, {\'U}lfar Erlingsson, Jernej Kos, and Dawn Song.
\newblock {The Secret Sharer: Evaluating and Testing Unintended Memorization in
  Neural Networks}.
\newblock In {\em {USENIX Security Symposium (USENIX Security)}}, pages
  267--284. USENIX, 2019.

\bibitem{CCGJY20}
Varun Chandrasekaran, Kamalika Chaudhuri, Irene Giacomelli, Somesh Jha, and
  Songbai Yan.
\newblock {Exploring Connections Between Active Learning and Model Extraction}.
\newblock In {\em {USENIX Security Symposium (USENIX Security)}}. USENIX, 2020.

\bibitem{CYZF20}
Dingfan Chen, Ning Yu, Yang Zhang, and Mario Fritz.
\newblock {GAN-Leaks: A Taxonomy of Membership Inference Attacks against GANs}.
\newblock In {\em {ACM SIGSAC Conference on Computer and Communications
  Security (CCS)}}. ACM, 2020.

\bibitem{CZWBHZ20}
Min Chen, Zhikun Zhang, Tianhao Wang, Michael Backes, Mathias Humbert, and Yang
  Zhang.
\newblock {When Machine Unlearning Jeopardizes Privacy}.
\newblock {\em {CoRR abs/2005.02205}}, 2020.

\bibitem{CXXLBKZ18}
Qingrong Chen, Chong Xiang, Minhui Xue, Bo~Li, Nikita Borisov, Dali Kaarfar,
  and Haojin Zhu.
\newblock {Differentially Private Data Generative Models}.
\newblock {\em {CoRR abs/1812.02274}}, 2018.

\bibitem{CNKMPAV17}
Yizheng Chen, Yacin Nadji, Athanasios Kountouras, Fabian Monrose, Roberto
  Perdisci, Manos Antonakakis, and Nikolaos Vasiloglou.
\newblock {Practical Attacks Against Graph-based Clustering}.
\newblock In {\em {ACM SIGSAC Conference on Computer and Communications
  Security (CCS)}}, pages 1125--1142. ACM, 2017.

\bibitem{CWSJ20}
Yizheng Chen, Shiqi Wang, Dongdong She, and Suman Jana.
\newblock {On Training Robust PDF Malware Classifiers}.
\newblock In {\em {USENIX Security Symposium (USENIX Security)}}. USENIX, 2020.

\bibitem{DLTHWZS18}
Hanjun Dai, Hui Li, Tian Tian, Xin Huang, Lin Wang, Jun Zhu, and Le~Song.
\newblock {Adversarial Attack on Graph Structured Data}.
\newblock In {\em {International Conference on Machine Learning (ICML)}}, pages
  1123--1132. PMLR, 2018.

\bibitem{DBV16}
Micha{\"e}l Defferrard, Xavier Bresson, and Pierre Vandergheynst.
\newblock {Convolutional Neural Networks on Graphs with Fast Localized Spectral
  Filtering}.
\newblock In {\em {Annual Conference on Neural Information Processing Systems
  (NIPS)}}, pages 3837--3845. NIPS, 2016.

\bibitem{DD03}
Paul~D. Dobson and Andrew~J. Doig.
\newblock {Distinguishing Enzyme Structures from Non-Enzymes without
  Alignments}.
\newblock {\em {Journal of Molecular Biology}}, 2003.

\bibitem{DJC15}
Yuxiao Dong, Reid~A. Johnson, and Nitesh~V. Chawla.
\newblock {Will This Paper Increase Your \emph{h}-index?: Scientific Impact
  Prediction}.
\newblock In {\em {ACM International Conference on Web Search and Data Mining
  (WSDM)}}, pages 149--158. ACM, 2015.

\bibitem{DJLBB20}
Vijay~Prakash Dwivedi, Chaitanya~K. Joshi, Thomas Laurent, Yoshua Bengio, and
  Xavier Bresson.
\newblock {Benchmarking Graph Neural Networks}.
\newblock {\em {CoRR abs/2003.00982}}, 2020.

\bibitem{EPBM20}
Federico Errica, Marco Podda, Davide Bacciu, and Alessio Micheli.
\newblock {A Fair Comparison of Graph Neural Networks for Graph
  Classification}.
\newblock In {\em {International Conference on Learning Representations
  (ICLR)}}, 2020.

\bibitem{FMLHZTY19}
Wenqi Fan, Yao Ma, Qing Li, Yuan He, Yihong~Eric Zhao, Jiliang Tang, and Dawei
  Yin.
\newblock {Graph Neural Networks for Social Recommendation}.
\newblock In {\em {The Web Conference (WWW)}}, pages 417--426. ACM, 2019.

\bibitem{FJR15}
Matt Fredrikson, Somesh Jha, and Thomas Ristenpart.
\newblock {Model Inversion Attacks that Exploit Confidence Information and
  Basic Countermeasures}.
\newblock In {\em {ACM SIGSAC Conference on Computer and Communications
  Security (CCS)}}, pages 1322--1333. ACM, 2015.

\bibitem{FLJLPR14}
Matt Fredrikson, Eric Lantz, Somesh Jha, Simon Lin, David Page, and Thomas
  Ristenpart.
\newblock {Privacy in Pharmacogenetics: An End-to-End Case Study of
  Personalized Warfarin Dosing}.
\newblock In {\em {USENIX Security Symposium (USENIX Security)}}, pages 17--32.
  USENIX, 2014.

\bibitem{GSRVD17}
Justin Gilmer, Samuel~S. Schoenholz, Patrick~F. Riley, Oriol Vinyals, and
  George~E. Dahl.
\newblock {Neural Message Passing for Quantum Chemistry}.
\newblock In {\em {International Conference on Machine Learning (ICML)}}, pages
  1263--1272. PMLR, 2017.

\bibitem{GL162}
Neil~Zhenqiang Gong and Bin Liu.
\newblock {You are Who You Know and How You Behave: Attribute Inference Attacks
  via Users' Social Friends and Behaviors}.
\newblock In {\em {USENIX Security Symposium (USENIX Security)}}, pages
  979--995. USENIX, 2016.

\bibitem{GTMHSSSS14}
Neil~Zhenqiang Gong, Ameet Talwalkar, Lester~W. Mackey, Ling Huang, Eui
  Chul~Richard Shin, Emil Stefanov, Elaine Shi, and Dawn Song.
\newblock {Joint Link Prediction and Attribute Inference Using a
  Social-Attribute Network}.
\newblock {\em {ACM Transactions on Intelligent Systems and Technology}}, 2014.

\bibitem{GL16}
Aditya Grover and Jure Leskovec.
\newblock {node2vec: Scalable Feature Learning for Networks}.
\newblock In {\em {ACM Conference on Knowledge Discovery and Data Mining
  (KDD)}}, pages 855--864. ACM, 2016.

\bibitem{HZHBTWB19}
Inken Hagestedt, Yang Zhang, Mathias Humbert, Pascal Berrang, Haixu Tang,
  XiaoFeng Wang, and Michael Backes.
\newblock {MBeacon: Privacy-Preserving Beacons for DNA Methylation Data}.
\newblock In {\em {Network and Distributed System Security Symposium (NDSS)}}.
  Internet Society, 2019.

\bibitem{HYL17}
William~L. Hamilton, Zhitao Ying, and Jure Leskovec.
\newblock {Inductive Representation Learning on Large Graphs}.
\newblock In {\em {Annual Conference on Neural Information Processing Systems
  (NIPS)}}, pages 1025--1035. NIPS, 2017.

\bibitem{HLMJ09}
Michael Hay, Chao Li, Gerome Miklau, and David~D. Jensen.
\newblock {Accurate Estimation of the Degree Distribution of Private Networks}.
\newblock In {\em {International Conference on Data Mining (ICDM)}}, pages
  169--178. IEEE, 2009.

\bibitem{HMDC19}
Jamie Hayes, Luca Melis, George Danezis, and Emiliano~De Cristofaro.
\newblock {LOGAN: Evaluating Privacy Leakage of Generative Models Using
  Generative Adversarial Networks}.
\newblock {\em {Symposium on Privacy Enhancing Technologies Symposium}}, 2019.

\bibitem{JCBKP20}
Matthew Jagielski, Nicholas Carlini, David Berthelot, Alex Kurakin, and Nicolas
  Papernot.
\newblock {High Accuracy and High Fidelity Extraction of Neural Networks}.
\newblock In {\em {USENIX Security Symposium (USENIX Security)}}. USENIX, 2020.

\bibitem{JG18}
Jinyuan Jia and Neil~Zhenqiang Gong.
\newblock {AttriGuard: A Practical Defense Against Attribute Inference Attacks
  via Adversarial Machine Learning}.
\newblock In {\em {USENIX Security Symposium (USENIX Security)}}, pages
  513--529. USENIX, 2018.

\bibitem{JSBZG19}
Jinyuan Jia, Ahmed Salem, Michael Backes, Yang Zhang, and Neil~Zhenqiang Gong.
\newblock {MemGuard: Defending against Black-Box Membership Inference Attacks
  via Adversarial Examples}.
\newblock In {\em {ACM SIGSAC Conference on Computer and Communications
  Security (CCS)}}, pages 259--274. ACM, 2019.

\bibitem{JWCG20}
Jinyuan Jia, Binghui Wang, Xiaoyu Cao, and Neil~Zhenqiang Gong.
\newblock {Certified Robustness of Community Detection against Adversarial
  Structural Perturbation via Randomized Smoothing}.
\newblock In {\em {The Web Conference (WWW)}}, pages 2718--2724. ACM, 2020.

\bibitem{JSMA19}
Mika Juuti, Sebastian Szyller, Samuel Marchal, and N.~Asokan.
\newblock {PRADA: Protecting Against DNN Model Stealing Attacks}.
\newblock In {\em {IEEE European Symposium on Security and Privacy (Euro
  S\&P)}}, pages 512--527. IEEE, 2019.

\bibitem{KW17}
Thomas~N. Kipf and Max Welling.
\newblock {Semi-Supervised Classification with Graph Convolutional Networks}.
\newblock In {\em {International Conference on Learning Representations
  (ICLR)}}, 2017.

\bibitem{LF20}
Klas Leino and Matt Fredrikson.
\newblock {Stolen Memories: Leveraging Model Memorization for Calibrated
  White-Box Membership Inference}.
\newblock In {\em {USENIX Security Symposium (USENIX Security)}}. USENIX, 2020.

\bibitem{LMXZ20}
Shaofeng Li, Shiqing Ma, Minhui Xue, and Benjamin Zi~Hao Zhao.
\newblock {Deep Learning Backdoors}.
\newblock {\em {CoRR abs/2007.08273}}, 2020.

\bibitem{LHZG19}
Zheng Li, Chengyu Hu, Yang Zhang, and Shanqing Guo.
\newblock {How to Prove Your Model Belongs to You: A Blind-Watermark based
  Framework to Protect Intellectual Property of DNN}.
\newblock In {\em {Annual Computer Security Applications Conference (ACSAC)}},
  pages 126--137. ACM, 2019.

\bibitem{LZ20}
Zheng Li and Yang Zhang.
\newblock {Label-Leaks: Membership Inference Attack with Label}.
\newblock {\em {CoRR abs/2007.15528}}, 2020.

\bibitem{LK07}
David Liben-Nowell and Jon Kleinberg.
\newblock {The Link-prediction Problem for Social Networks}.
\newblock {\em {Journal of the American Society for Information Science and
  Technology}}, 2007.

\bibitem{MSCS19}
Luca Melis, Congzheng Song, Emiliano~De Cristofaro, and Vitaly Shmatikov.
\newblock {Exploiting Unintended Feature Leakage in Collaborative Learning}.
\newblock In {\em {IEEE Symposium on Security and Privacy (S\&P)}}, pages
  497--512. IEEE, 2019.

\bibitem{NSH18}
Milad Nasr, Reza Shokri, and Amir Houmansadr.
\newblock {Machine Learning with Membership Privacy using Adversarial
  Regularization}.
\newblock In {\em {ACM SIGSAC Conference on Computer and Communications
  Security (CCS)}}, pages 634--646. ACM, 2018.

\bibitem{NSH19}
Milad Nasr, Reza Shokri, and Amir Houmansadr.
\newblock {Comprehensive Privacy Analysis of Deep Learning: Passive and Active
  White-box Inference Attacks against Centralized and Federated Learning}.
\newblock In {\em {IEEE Symposium on Security and Privacy (S\&P)}}, pages
  1021--1035. IEEE, 2019.

\bibitem{OSF19}
Tribhuvanesh Orekondy, Bernt Schiele, and Mario Fritz.
\newblock {Knockoff Nets: Stealing Functionality of Black-Box Models}.
\newblock In {\em {IEEE Conference on Computer Vision and Pattern Recognition
  (CVPR)}}, pages 4954--4963. IEEE, 2019.

\bibitem{OSF20}
Tribhuvanesh Orekondy, Bernt Schiele, and Mario Fritz.
\newblock {Prediction Poisoning: Towards Defenses Against DNN Model Stealing
  Attacks}.
\newblock In {\em {International Conference on Learning Representations
  (ICLR)}}, 2020.

\bibitem{PZ172}
Jun Pang and Yang Zhang.
\newblock {DeepCity: A Feature Learning Framework for Mining Location
  Check-Ins}.
\newblock In {\em {International Conference on Weblogs and Social Media
  (ICWSM)}}, pages 652--655. AAAI, 2017.

\bibitem{PZ17}
Jun Pang and Yang Zhang.
\newblock {Quantifying Location Sociality}.
\newblock In {\em {ACM Conference on Hypertext and Social Media (HT)}}, pages
  145--154. ACM, 2017.

\bibitem{PMSW18}
Nicolas Papernot, Patrick McDaniel, Arunesh Sinha, and Michael Wellman.
\newblock {SoK: Towards the Science of Security and Privacy in Machine
  Learning}.
\newblock In {\em {IEEE European Symposium on Security and Privacy (Euro
  S\&P)}}, pages 399--414. IEEE, 2018.

\bibitem{PMGJCS17}
Nicolas Papernot, Patrick~D. McDaniel, Ian Goodfellow, Somesh Jha, Z.~Berkay
  Celik, and Ananthram Swami.
\newblock {Practical Black-Box Attacks Against Machine Learning}.
\newblock In {\em {ACM Asia Conference on Computer and Communications Security
  (ASIACCS)}}, pages 506--519. ACM, 2017.

\bibitem{QMR19}
Erwin Quiring, Alwin Maier, and Konrad Rieck.
\newblock {Misleading Authorship Attribution of Source Code using Adversarial
  Learning}.
\newblock In {\em {USENIX Security Symposium (USENIX Security)}}, pages
  479--496. USENIX, 2019.

\bibitem{RB08}
Kaspar Riesen and Horst Bunke.
\newblock {\em {Structural, Syntactic, and Statistical Pattern Recognition}}.
\newblock Springer, 2008.

\bibitem{SBBFZ20}
Ahmed Salem, Apratim Bhattacharya, Michael Backes, Mario Fritz, and Yang Zhang.
\newblock {Updates-Leak: Data Set Inference and Reconstruction Attacks in
  Online Learning}.
\newblock In {\em {USENIX Security Symposium (USENIX Security)}}, pages
  1291--1308. USENIX, 2020.

\bibitem{SWBMZ20}
Ahmed Salem, Rui Wen, Michael Backes, Shiqing Ma, and Yang Zhang.
\newblock {Dynamic Backdoor Attacks Against Machine Learning Models}.
\newblock {\em {CoRR abs/2003.03675}}, 2020.

\bibitem{SZHBFB19}
Ahmed Salem, Yang Zhang, Mathias Humbert, Pascal Berrang, Mario Fritz, and
  Michael Backes.
\newblock {ML-Leaks: Model and Data Independent Membership Inference Attacks
  and Defenses on Machine Learning Models}.
\newblock In {\em {Network and Distributed System Security Symposium (NDSS)}}.
  Internet Society, 2019.

\bibitem{SHNSSDG18}
Ali Shafahi, W~Ronny Huang, Mahyar Najibi, Octavian Suciu, Christoph Studer,
  Tudor Dumitras, and Tom Goldstein.
\newblock {Poison Frogs! Targeted Clean-Label Poisoning Attacks on Neural
  Networks}.
\newblock In {\em {Annual Conference on Neural Information Processing Systems
  (NeurIPS)}}, pages 6103--6113. NeurIPS, 2018.

\bibitem{SSSS17}
Reza Shokri, Marco Stronati, Congzheng Song, and Vitaly Shmatikov.
\newblock {Membership Inference Attacks Against Machine Learning Models}.
\newblock In {\em {IEEE Symposium on Security and Privacy (S\&P)}}, pages
  3--18. IEEE, 2017.

\bibitem{SS19}
Congzheng Song and Vitaly Shmatikov.
\newblock {Auditing Data Provenance in Text-Generation Models}.
\newblock In {\em {ACM Conference on Knowledge Discovery and Data Mining
  (KDD)}}, pages 196--206. ACM, 2019.

\bibitem{SS202}
Congzheng Song and Reza Shokri.
\newblock {Robust Membership Encoding: Inference Attacks and Copyright
  Protection for Deep Learning}.
\newblock In {\em {ACM Asia Conference on Computer and Communications Security
  (ASIACCS)}}. ACM, 2020.

\bibitem{SOW03}
Jeffrey Sutherland, Lee O'Brien, and Donald Weaver.
\newblock {SplineFitting with a Genetic Algorithm: A Method for Developing
  Classification Structure Activity Relationships}.
\newblock {\em {Journal of Chemical Information and Computer Sciences}}, 2003.

\bibitem{TZJRR16}
Florian Tram{\`e}r, Fan Zhang, Ari Juels, Michael~K. Reiter, and Thomas
  Ristenpart.
\newblock {Stealing Machine Learning Models via Prediction APIs}.
\newblock In {\em {USENIX Security Symposium (USENIX Security)}}, pages
  601--618. USENIX, 2016.

\bibitem{MH08}
Laurens van~der Maaten and Geoffrey Hinton.
\newblock {Visualizing Data using {t-SNE}}.
\newblock {\em {Journal of Machine Learning Research}}, 2008.

\bibitem{VCCRLB18}
Petar Velickovic, Guillem Cucurull, Arantxa Casanova, Adriana Romero, Pietro
  Li{\`o}, and Yoshua Bengio.
\newblock {Graph Attention Networks}.
\newblock In {\em {International Conference on Learning Representations
  (ICLR)}}, 2018.

\bibitem{WG18}
Binghui Wang and Neil~Zhenqiang Gong.
\newblock {Stealing Hyperparameters in Machine Learning}.
\newblock In {\em {IEEE Symposium on Security and Privacy (S\&P)}}, pages
  36--52. IEEE, 2018.

\bibitem{WG19}
Binghui Wang and Neil~Zhenqiang Gong.
\newblock {Attacking Graph-based Classification via Manipulating the Graph
  Structure}.
\newblock In {\em {ACM SIGSAC Conference on Computer and Communications
  Security (CCS)}}, pages 2023--2040. ACM, 2019.

\bibitem{WJG19}
Binghui Wang, Jinyuan Jia, and Neil~Zhenqiang Gong.
\newblock {Graph-based Security and Privacy Analytics via Collective
  Classification with Joint Weight Learning and Propagation}.
\newblock In {\em {Network and Distributed System Security Symposium (NDSS)}}.
  Internet Society, 2019.

\bibitem{WWTDLZ19}
Huijun Wu, Chen Wang, Yuriy Tyshetskiy, Andrew Docherty, Kai Lu, and Liming
  Zhu.
\newblock {Adversarial Examples for Graph Data: Deep Insights into Attack and
  Defense}.
\newblock In {\em {International Joint Conferences on Artifical Intelligence
  (IJCAI)}}, pages 4816--4823. IJCAI, 2019.

\bibitem{YGFJ18}
Samuel Yeom, Irene Giacomelli, Matt Fredrikson, and Somesh Jha.
\newblock {Privacy Risk in Machine Learning: Analyzing the Connection to
  Overfitting}.
\newblock In {\em {IEEE Computer Security Foundations Symposium (CSF)}}, pages
  268--282. IEEE, 2018.

\bibitem{ZCPSX15}
Jun Zhang, Graham Cormode, Cecilia~M. Procopiuc, Divesh Srivastava, and Xiaokui
  Xiao.
\newblock {Private Release of Graph Statistics using Ladder Functions}.
\newblock In {\em {ACM SIGMOD International Conference on Management of Data
  (SIGMOD)}}, pages 731--745. ACM, 2015.

\bibitem{Z19}
Yang Zhang.
\newblock {Language in Our Time: An Empirical Analysis of Hashtags}.
\newblock In {\em {The Web Conference (WWW)}}, pages 2378--2389. ACM, 2019.

\bibitem{ZHSMVB20}
Yang Zhang, Mathias Humbert, Bartlomiej Surma, Praveen Manoharan, Jilles
  Vreeken, and Michael Backes.
\newblock {Towards Plausible Graph Anonymization}.
\newblock In {\em {Network and Distributed System Security Symposium (NDSS)}}.
  Internet Society, 2020.

\bibitem{ZJWG20}
Zaixi Zhang, Jinyuan Jia, Binghui Wang, and Neil~Zhenqiang Gong.
\newblock {Backdoor Attacks to Graph Neural Networks}.
\newblock {\em {CoRR abs/2006.11165}}, 2020.

\bibitem{ZZCZ19}
Dingyuan Zhu, Ziwei Zhang, Peng Cui, and Wenwu Zhu.
\newblock {Robust Graph Convolutional Networks Against Adversarial Attacks}.
\newblock In {\em {ACM Conference on Knowledge Discovery and Data Mining
  (KDD)}}, pages 1399--1407. ACM, 2019.

\bibitem{ZAG18}
Daniel Z{\"u}gner, Amir Akbarnejad, and Stephan G{\"u}nnemann.
\newblock {Adversarial Attacks on Neural Networks for Graph Data}.
\newblock In {\em {ACM Conference on Knowledge Discovery and Data Mining
  (KDD)}}, pages 2847--2856. ACM, 2018.

\bibitem{ZG19}
Daniel Z{\"u}gner and Stephan G{\"u}nnemann.
\newblock {Adversarial Attacks on Graph Neural Networks via Meta Learning}.
\newblock In {\em {International Conference on Learning Representations
  (ICLR)}}, 2019.

\bibitem{ZG192}
Daniel Z{\"u}gner and Stephan G{\"u}nnemann.
\newblock {Certifiable Robustness and Robust Training for Graph Convolutional
  Networks}.
\newblock In {\em {ACM Conference on Knowledge Discovery and Data Mining
  (KDD)}}, pages 246--256. ACM, 2019.

\end{thebibliography}
% ----------------------------------------------------

% ----------------------------------------------------
\appendix
\section{Appendix}
\label{section:Appendix}
% ----------------------------------------------------

\begin{table}[!h]
\centering
\caption{
Distance metrics, $\TargetModel_i(u)$ represents the $i$-th component of $\TargetModel(u)$. 
Note that these metrics can be applied to nodes' attributes as well.
}
\label{table:distance}
\footnotesize
\begin{tabular}{l|c}
\toprule
Metrics & Definition \\
\midrule
Cosine & $1 - \dfrac{\TargetModel(u)\cdot \TargetModel(v)}{\left\|\TargetModel(u)\right\|_2\left\|\TargetModel(v)\right\|_2}$ \\
Euclidean & $\left\|\TargetModel(u) - \TargetModel(v)\right\|_2$ \\
Correlation & $1-\dfrac{(\TargetModel(u)-\overline{\TargetModel(u)}) \cdot(\TargetModel(v)-\overline{\TargetModel(v)})}{\|(\TargetModel(u)-\overline{\TargetModel(u)})\|_{2}\|(\TargetModel(v)-\overline{\TargetModel(v)})\|_{2}}$ \\
Chebyshev & $\max _{i}\left|\TargetModel_i(u)-\TargetModel_i(v)\right|$ \\
Braycurtis & $\dfrac{\sum\left|\TargetModel_i(u)-\TargetModel_i(v)\right|} {\sum\left|\TargetModel_i(u)+\TargetModel_i(v)\right|}$ \\
Manhattan & $\sum_{i}\left|\TargetModel_i(u)-\TargetModel_i(v)\right|$ \\
Canberra & $\sum_{i} \dfrac{\left|\TargetModel_i(u)-\TargetModel_i(v)\right|}{\left|\TargetModel_i(u)\right|+\left|\TargetModel_i(v)\right|}$ \\
Sqeuclidean & $\left\|\TargetModel(u) - \TargetModel(v)\right\|_2^2$ \\
\bottomrule
\end{tabular}
\end{table}

\begin{table}[!h]
\centering
\caption{
Pairwise vector operations, $\TargetModel_i(u)$ represents the $i$-th component of $\TargetModel(u)$.
Note that these operations can be applied to nodes' attributes and entropies summarized from posteriors as well.
}
\label{table:operator}
\footnotesize
\begin{tabular}{l|c|l|c}
\toprule
Operator & Definition & Operator & Definition\\
\midrule
Average & $\dfrac{\TargetModel_i(u)+\TargetModel_i(v)}{2}$ & Weighted-L1 & $|\TargetModel_i(u) - \TargetModel_i(v)|$\\
Hadamard & $\TargetModel_i(u) \cdot \TargetModel_i(v)$ & Weighted-L2 & $|\TargetModel_i(u) - \TargetModel_i(v)|^2$\\
\bottomrule
\end{tabular}
\end{table}

\begin{table}[!ht]
\centering
\caption{
Prediction results for Attack-0 on all the 8 datasets with Correlation distance.
}
\label{table:attack_0_prf}
\footnotesize
\setlength{\tabcolsep}{0.6em}
\begin{tabular}{l|cccc}
\toprule
Dataset & Precision & Recall & F1-Score & AUC\\
\midrule
AIDS            & 0.524  & 0.996 & 0.687 & 0.691\\
COX2            & 0.523  & 0.987 & 0.684 & 0.867\\
DHFR            & 0.555  & 0.977 & 0.708 & 0.765\\
ENZYMES         & 0.501  & 1.000 & 0.667 & 0.630\\
PROTEINS\_full  & 0.540  & 0.998 & 0.701 & 0.815\\
Citeseer        & 0.788  & 0.991 & 0.878 & 0.959\\
Cora            & 0.777  & 0.966 & 0.861 & 0.929\\
Pubmed          & 0.691  & 0.965 & 0.806 & 0.874\\
\bottomrule
\end{tabular}
\end{table}

\begin{table}[!ht]
\centering
\caption{
Average AUC with standard deviation for Attack-1 with different GCN structures on all the 8 datasets.
Results with respect to the best performing shadow dataset are reported.
}
\label{table:attack1_different_layer}
\footnotesize
\setlength{\tabcolsep}{0.3em}
\begin{tabular}{l|l|c}
\toprule
Dataset & Shadow Dataset & AUC \\
\midrule
AIDS & PROTEINS\_full & 0.729 $\pm$ 0.013 \\
COX2 & Citeseer & 0.760 $\pm$ 0.026 \\
DHFR & COX2 & 0.792 $\pm$ 0.005 \\
ENZYMES & AIDS & 0.732 $\pm$ 0.009 \\
PROTEINS\_full & COX2 & 0.808 $\pm$ 0.034  \\
Citeseer & Cora & 0.924 $\pm$ 0.006  \\
Cora & Citeseer & 0.916 $\pm$ 0.002 \\
Pubmed & Citeseer & 0.840 $\pm$ 0.001  \\
\bottomrule
\end{tabular}
\end{table}

\begin{figure*}[!ht]
\centering
\includegraphics[width=2\columnwidth]{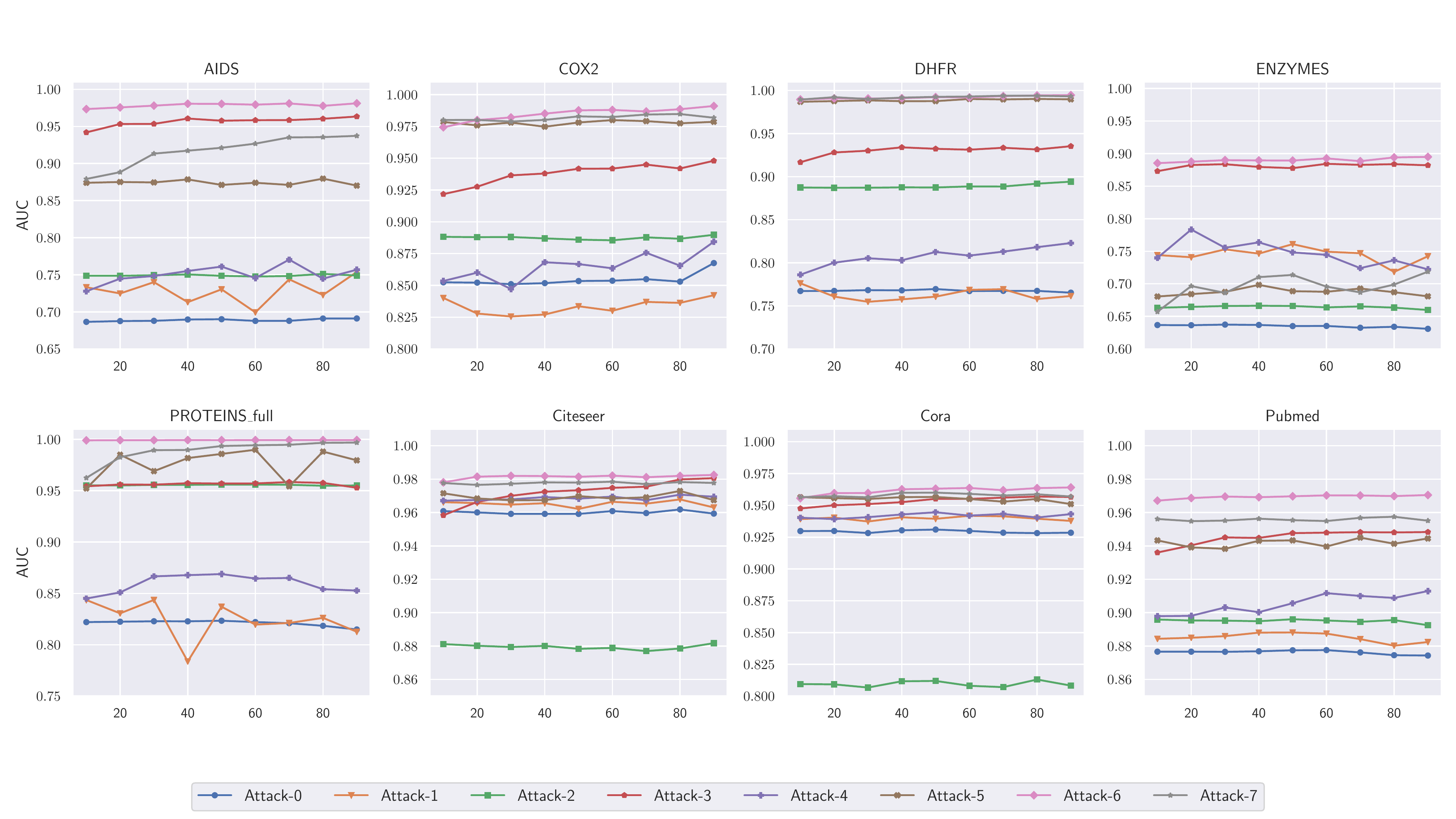}
\caption{
The relationship between the ratio of attack training dataset in the attack dataset and the attacks' AUC scores on all the 8 datasets.
The x-axis represents the ratio and the y-axis represents the AUC score.
}
\label{figure:attack_different_ratio}
\end{figure*} 

\begin{table}[!ht]
\centering
\caption{
Average Precision and Recall with standard deviation for Attack-1. 
Results with respect to the best performing shadow dataset are reported.
}
\label{table:attack1_pr}
\footnotesize
\setlength{\tabcolsep}{0.3em}
\begin{tabular}{l|l|cc}
\toprule
Dataset & Shadow Dataset & Precision & Recall\\
\midrule
AIDS & ENZYMES & 0.725 $\pm$ 0.044 & 0.505 $\pm$ 0.110 \\
COX2 & PROTEINS\_full & 0.828 $\pm$ 0.013 & 0.686 $\pm$ 0.100 \\
DHFR & COX2 & 0.691 $\pm$ 0.015 & 0.704 $\pm$ 0.022 \\
ENZYMES & AIDS & 0.639 $\pm$ 0.023 & 0.615 $\pm$ 0.046 \\
PROTEINS\_full & Citeseer & 0.750 $\pm$ 0.022 & 0.800 $\pm$ 0.055 \\
Citeseer & Cora & 0.871 $\pm$ 0.005 & 0.958 $\pm$ 0.005 \\
Cora & Citeseer & 0.854 $\pm$ 0.003 & 0.883 $\pm$ 0.008 \\
Pubmed & Cora & 0.765 $\pm$ 0.009 & 0.897 $\pm$ 0.012 \\
\bottomrule
\end{tabular}
\end{table}

\begin{table}[!ht]
\centering
\caption{
Average Precision and Recall with standard deviation for Attack-3.
}
\label{table:attack3_pr}
\footnotesize
\setlength{\tabcolsep}{0.3em}
\begin{tabular}{l|cc}
\toprule
Dataset  & Precision & Recall\\
\midrule
AIDS &  0.874 $\pm$ 0.006 & 0.966 $\pm$ 0.005 \\
COX2 &  0.846 $\pm$ 0.004 & 0.922 $\pm$ 0.005 \\
DHFR &  0.847 $\pm$ 0.007 & 0.877 $\pm$ 0.009 \\
ENZYMES &  0.761 $\pm$ 0.003 & 0.871 $\pm$ 0.004 \\
PROTEINS\_full &  0.856 $\pm$ 0.006 & 0.943 $\pm$ 0.004 \\
Citeseer &  0.895 $\pm$ 0.003 & 0.946 $\pm$ 0.005 \\
Cora &  0.858 $\pm$ 0.002 & 0.917 $\pm$ 0.008 \\
Pubmed &  0.869 $\pm$ 0.008 & 0.892 $\pm$ 0.014 \\
\bottomrule
\end{tabular}
\end{table}

\begin{table}[!ht]
\centering
\caption{
Average Precision and Recall with standard deviation for Attack-4. 
Results with respect to the best performing shadow dataset are reported.
}
\label{table:attack4_pr}
\footnotesize
\setlength{\tabcolsep}{0.3em}
\begin{tabular}{l|l|cc}
\toprule
Dataset & Shadow Dataset & Precision & Recall\\
\midrule
AIDS & DHFR & 0.688 $\pm$ 0.013 & 0.628 $\pm$ 0.046 \\
COX2 & DHFR & 0.787 $\pm$ 0.009 & 0.835 $\pm$ 0.033 \\
DHFR & COX2 & 0.726 $\pm$ 0.008 & 0.793 $\pm$ 0.015 \\
ENZYMES & AIDS & 0.637 $\pm$ 0.025 & 0.683 $\pm$ 0.041 \\
PROTEINS\_full & Pubmed & 0.686 $\pm$ 0.045 & 0.955 $\pm$ 0.020 \\
Citeseer & Cora & 0.874 $\pm$ 0.004 & 0.956 $\pm$ 0.004 \\
Cora & Citeseer & 0.854 $\pm$ 0.002 & 0.896 $\pm$ 0.004 \\
Pubmed & Citeseer & 0.790 $\pm$ 0.009 & 0.877 $\pm$ 0.012 \\
\bottomrule
\end{tabular}
\end{table}

\begin{table}[!ht]
\centering
\caption{
Average Precision and Recall with standard deviation for Attack-5. 
Results with respect to the best performing shadow dataset are reported.
}
\label{table:attack5_pr}
\footnotesize
\setlength{\tabcolsep}{0.3em}
\begin{tabular}{l|l|cc}
\toprule
Dataset & Shadow Dataset & Precision & Recall\\
\midrule
AIDS & PROTEINS\_full & 0.854 $\pm$ 0.003 & 0.663 $\pm$ 0.005 \\
COX2 & DHFR & 0.941 $\pm$ 0.004 & 0.923 $\pm$ 0.022 \\
DHFR & COX2 & 0.973 $\pm$ 0.004 & 0.942 $\pm$ 0.025 \\
ENZYMES & Citeseer & 0.608 $\pm$ 0.005 & 0.675 $\pm$ 0.013 \\
PROTEINS\_full & COX2 & 0.996 $\pm$ 0.003 & 0.061 $\pm$ 0.055 \\
Citeseer & Cora & 0.888 $\pm$ 0.006 & 0.885 $\pm$ 0.005 \\
Cora & Citeseer & 0.867 $\pm$ 0.006 & 0.892 $\pm$ 0.009 \\
Pubmed & Cora & 0.824 $\pm$ 0.010 & 0.913 $\pm$ 0.014 \\
\bottomrule
\end{tabular}
\end{table}

\begin{table}[!ht]
\centering
\caption{
Average Precision and Recall with standard deviation for Attack-6.
}
\label{table:attack6_pr}
\footnotesize
\setlength{\tabcolsep}{0.3em}
\begin{tabular}{l|cc}
\toprule
Dataset  & Precision & Recall\\
\midrule
AIDS &  0.907 $\pm$ 0.002 & 0.986 $\pm$ 0.002 \\
COX2 &  0.935 $\pm$ 0.004 & 0.994 $\pm$ 0.001 \\
DHFR &  0.972 $\pm$ 0.001 & 0.995 $\pm$ 0.002 \\
ENZYMES &  0.770 $\pm$ 0.004 & 0.886 $\pm$ 0.009 \\
PROTEINS\_full &  0.988 $\pm$ 0.002 & 0.998 $\pm$ 0.001 \\
Citeseer &  0.900 $\pm$ 0.008 & 0.933 $\pm$ 0.006 \\
Cora &  0.878 $\pm$ 0.003 & 0.930 $\pm$ 0.003 \\
Pubmed &  0.903 $\pm$ 0.004 & 0.920 $\pm$ 0.003 \\
\bottomrule
\end{tabular}
\end{table}

\begin{table}[!ht]
\centering
\caption{
Average Precision and Recall with standard deviation for Attack-7. 
Results with respect to the best performing shadow dataset are reported.
}
\label{table:attack7_pr}
\footnotesize
\setlength{\tabcolsep}{0.3em}
\begin{tabular}{l|l|cc}
\toprule
Dataset & Shadow Dataset & Precision & Recall\\
\midrule
AIDS & COX2 & 0.870 $\pm$ 0.003 & 0.781 $\pm$ 0.013 \\
COX2 & DHFR & 0.941 $\pm$ 0.004 & 0.966 $\pm$ 0.009 \\
DHFR & COX2 & 0.972 $\pm$ 0.002 & 0.994 $\pm$ 0.005 \\
ENZYMES & AIDS & 0.617 $\pm$ 0.012 & 0.693 $\pm$ 0.036 \\
PROTEINS\_full & ENZYMES & 0.955 $\pm$ 0.004 & 0.974 $\pm$ 0.010 \\
Citeseer & Cora & 0.898 $\pm$ 0.003 & 0.913 $\pm$ 0.008 \\
Cora & Citeseer & 0.874 $\pm$ 0.004 & 0.911 $\pm$ 0.005 \\
Pubmed & Citeseer & 0.881 $\pm$ 0.006 & 0.901 $\pm$ 0.010 \\
\bottomrule
\end{tabular}
\end{table}

% ----------------------------------------------------
\end{document}